\documentclass[aps,prd,showpacs,showkeys,nofootinbib,eqsecnum]{revtex4}
\usepackage{epsfig}% Include figure files
\input{epsf}

\begin{document}
\def\be{\begin{equation}}
\def\ee{\end{equation}}
\def\beq{\begin{equation}}
\def\eeq{\end{equation}}
\def\eq{\beq\eeq}
\def\bea{\begin{eqnarray}}
\def\eea{\end{eqnarray}}
\def\ds{\displaystyle}
\newcommand{\nn}{\nonumber}
\newcommand{\lsim}{\raisebox{-0.13cm}{~\shortstack{$<$ \\[-0.07cm] $\sim$}}~}
\newcommand{\Lam}{\Lambda}
\newcommand{\eps}{\epsilon} 
\newcommand{\MS}{\overline{MS}} 
\newcommand{\qq}{\langle \bar q q \rangle} 
\newcommand{\CSB}{\raisebox{.8mm}{$\chi$}SB }
\preprint{PM/01-68}
\preprint{hep-th/0205133}
\title{ (Borel) convergence of the  variationally improved mass
 expansion \\ 
and the $O(N)$ Gross-Neveu model mass gap}
\author{J.-L. Kneur and D. Reynaud}
\affiliation{Physique Math\'ematique et Th\'eorique, UMR-5825-CNRS, \\
Universit\'e Montpellier II, F--34095 Montpellier Cedex 5, France.}
%\date{\today}

\begin{abstract}
We reconsider in some detail a construction 
 allowing (Borel) convergence of  
an alternative perturbative expansion,
for specific physical quantities of asymptotically free
models. The usual perturbative expansions     
(with an explicit mass dependence) 
are transmuted into expansions in $1/F$, where 
$F \sim 1/g(m)$ for $m \gg \Lambda$ while $F \sim (m/\Lambda)^\alpha$
for $m \lsim \Lambda$, 
$\Lambda$ being the basic scale and $\alpha$ given by
renormalization group coefficients. (Borel) convergence holds 
in a range of $F$ which corresponds
to reach unambiguously the strong coupling infrared regime near
$m\to 0$, which can define certain ``non-perturbative" quantities, such
as the mass gap, from a resummation of this alternative expansion. 
Convergence properties can be further improved,  
when combined with $\delta$ expansion (variationally improved 
perturbation) methods. We
illustrate these results by re-evaluating, from purely perturbative
informations, the $O(N)$ Gross-Neveu model mass gap, known for arbitrary $N$
from exact S matrix results.  Comparing different levels of
approximations that can be defined within our framework, 
we find reasonable agreement with the exact result.
\end{abstract}
\maketitle
\setcounter{page}{1}
\section{Introduction}
In many quantum field models,
non-perturbative results may be obtained from 
the $1/N$ expansion~\cite{vectorN,gaugeN}, which at leading orders
resums only a certain class of graphs of the original perturbation series
in the coupling. In parallel, for a given perturbative series 
there are more direct efficient summation techniques,
like the Borel\cite{Borel,renormalons} or Pad\'e\cite{pade} methods, 
as well as generalizations combining the latter two\cite{aoBsum,PadBor}.
Typically, the Borel method
is useful even for non Borel-summable perturbative expansions, as for
instance in QCD, since it gives interesting informations on the 
incompleteness of the pure perturbation theory, and the
necessary additional
non-perturbative (power corrections) contributions
to a given physical quantity.  Also,
a rather different modification of the usual perturbation theory, known as
delta-expansion (DE) or ``variationally improved perturbation" 
(VIP)\cite{odm,delta,pms}, is based
on a reorganization of the interaction Lagrangian such that it depends on
arbitrary adjustable parameters, to be fixed by some optimization prescription.
In $D=1$ field theories, the quantum mechanical anharmonic
oscillator typically, DE-VIP is in fact equivalent\cite{deltaconv} to the
``order-dependent mapping" (ODM) resummation method\cite{odm}, and
optimization is  equivalent
to a rescaling of the adjustable oscillator
mass with perturbative order, which can essentially suppress the
factorial large order behaviour of ordinary perturbative coefficients. 
This appropriate rescaling of a trial mass parameter was proven
to give a rigorously convergent series\cite{deltaconv,deltac}
e.g. for the oscillator energy levels\cite{ao} and related quantities.

In the present paper we reconsider yet
another alternative expansion, proposed some
time ago\cite{gn2}--\cite{qcd2}, which is close to (and partly inspired by) the
DE-VIP idea, but particularly suited to apply generically to arbitrary
higher dimensional (renormalizable) models. 
Rather similarly with the $1/N$ expansion, it starts with a specific
approximation  but can include, at least in
principle, the full information from the Lagrangian in a systematical way.
The basic construction exploits a physically motivated renormalization
group (RG) ``self-consistent mass" solution, which resums RG dependence to all
orders (at least in specific renormalization schemes). Moreover it
provides a non-perturbative information, encoded
in the infrared properties of an implicit function $F(\hat m)$, which 
may be viewed as a generalization of the logarithm,  
and occurring as the exact solution of the above mentioned
RG equation with the self-consistent mass boundary condition. 
At first RG order, $F(\hat m) \equiv \ln (\hat m/\Lam) -A \ln F(\hat m)$
is essentially the Lambert function\cite{Lambert},
where $A$ depends in a simple way on the  mass and
coupling RG coefficients ($\hat m$ is the renormalization  
scale-invariant Lagrangian mass and $\Lambda$ the
basic RG scale). Unlike the logarithm, however, $F(\hat m)$ has a power
expansion behaviour in $(\hat m/\Lambda)^{1/A}$ in the infrared, for 
$|\hat m| \le \hat m_c \lsim \Lam$, while it
matches the ordinary perturbative effective coupling, $ F  \sim  \ln
(\hat m/\Lam)  \sim 1/g(\hat m) $, for the  short distance $m \gg \Lam$
perturbative regime. As we shall argue, $F$ thus defines a rigorous (analytic)
bridge between the  usual perturbative short distance regime,  
and the strongly coupled, non-perturbative, 
massless (chiral) limit corresponding to $\hat m \lsim \Lam$, or 
$\hat m \ll \Lam$. This
transmutes the ordinary expansion (in the coupling $g$) of
physical (on-shell) Green functions, depending explicitly on a single mass
$m$, into a $1/F$ expansion, or equivalently a (mass) power expansion
in $(\hat m/\Lam)^{1/A}$ for sufficiently small $\hat m$. 
The main idea is to use those properties of $F$, 
in asymptotically free theories (AFT), to  infer a non trivial
mass gap $M(\hat m \to 0) \ne 0$\cite{gn2} in the deep infrared (strongly
coupled) regime, equivalently here the massless limit $\hat m \to 0$. More
generally other physical quantities can be derived similarly,
for instance the quark condensate $\qq (m \to 0)$, one of the order parameter
 of chiral symmetry breaking
(\CSB) in QCD\cite{qcd1,qcd2}), from their known perturbative expression for
$\hat m \gg \Lam$. At this stage, it is important to remark that our
construction is not by itself a proof of dynamical (chiral) symmetry breaking,
and applies indeed independently of whether chiral symmetry is (dynamically)
broken or not: it introduces rather an explicit chiral symmetry breaking mass
$\hat m$ in the Lagrangian, in such a way that the properties of $F(\hat m)$
encode a non trivial $M/\Lam (\hat m \to 0)$ ratio, smoothly extrapolated from
the massive case $M(\hat m \neq 0)$. This is to be simply viewed as a
generalization, for $m \neq 0$, of dimensional transmutation\cite{ColWei}. \\ 
Unfortunately,  such an extrapolation to $\hat m\to 0$ is well-defined
as far as the pure RG dependence of the relevant physical
quantities is concerned, while it turns out to be badly afflicted when
considering for the latter their complete perturbative
(non-RG-dependent)  expansion coefficients with their expected behaviour at
large orders. As is well-known, in most models the leading large order
behaviour of purely perturbative coefficients exhibit same signs (thus non
Borel summable) factorial divergences (the infrared renormalon
singularities\cite{thooft,renormalons}). 
The standard and seemingly unavoidable interpretation 
is that it implies large perturbative
ambiguities, e.g. of ${\cal O}(\Lam)$ for the (pole) mass gap, reflecting 
incompleteness of purely perturbative expansions and the necessity of adding
non-perturbative power corrections. However, our alternative
expansion  can be smoothly extrapolated down to small, and even {\em negative}
(or more generally complex) values of the expansion parameter $1/F$, thanks
again to the properties of $F$, in such a way that the corresponding
perturbative series can be Borel summable\cite{KRlet}. More precidely, in the
simplest situation (corresponding to the RG
parameter particular value $A=1$), there is one branch of $F$ such that $F <0$,
which simply produces the required sign-alternation in the perturbative
coefficients $\sim F^{-n}$. In this particularly simple case 
$A=1$, the range where $F<0$ happens to correspond also to  $\hat m
<0$. This is not a problem in principle, since in relativistically invariant
theories the absolute sign of the  Lagrangian mass term is irrelevant to
physical quantities, moreover in our context the physically relevant results
are in the massless (chiral symmetric) limit anyway. Thus taking $Re[F]<0$ 
($Re[\hat m] \lsim 0$) simply corresponds physically to reach a strongly
coupled regime near the massless Lagrangian limit, but without the usual
ambiguities from renormalons. More generally, i.e. for an arbitrary AFT in an
arbitrary scheme, as we shall examine there exist (complex) branches of
$F$ near the relevant massless limit, which is sufficient to ensure Borel
summability, the Borel singularities being moved away from the real axis.
Those different branches of $F$ may correspond either to
$\hat m>0$ or to $\hat m <0$. We shall argue that the actual
physical result is indeed independent of the branch on which the massless limit
is reached, though the (unphysical)
perturbative expansions have obviously different Borel summation properties
depending on the branch of $F$ considered.\\   
Independently of these Borel
convergence properties of the $1/F$-series, we can also consider, in a second
stage, an appropriate version of the (order--dependently rescaled)
``variationally improved" perturbation (DE-VIP), to be performed on the series
in $1/F(\hat m/\Lam)$, essentially replacing the true physical mass  by an
arbitrary adjustable, trial mass parameter. 
This produces a renormalization scheme (RS) dependent
factorial  {\em damping} of the original perturbative
coefficients at large orders, similarly
to the oscillator
case\cite{deltaconv,deltac}. Now here, the damping appears
insufficient to make the DE-VIP series readily convergent for arbitrary $m$, 
but can further improve\cite{KRlet} the Borel
convergence properties, which are also obtained typically for $Re[F] <0$.\\ 
The basics of our construction was defined
before\cite{gn2}, and some of its
phenomenological applications    
explored, to some extent, in QCD~\cite{qcd1,qcd2}. 
However, those previous numerical results 
were based on rather 
ad hoc approximations, either by
constructing approximants only based on the lowest orders
of the perturbative expansion,
and by optimization with respect to the renormalization
scale and/or scheme\cite{qcd2}. In particular, it ignored completely
the above mentioned ambiguities due to the large order, factorial behaviour
of perturbative expansions. We
are thus mainly concerned in the present paper to provide
a more concrete illustration of the formal Borel convergence
results obtained in \cite{KRlet}, by considering the mass gap of the
$O(N)$ Gross-Neveu (GN) model\cite{GN,gnnext}, known 
for arbitrary $N$ values from exact S matrix
results\cite{exactS} and Thermodynamic Bethe Ansatz\cite{FNW} similarly 
applicable\cite{TBArev}
in many other integrable 2-D models.
Those exact results 
serve as a test of our method, which is based in contrast only on the
perturbative information, thus applying a priori to any
other (asymptotically free) renormalizable models, e.g. in four dimensions, for
which there are obviously no exact S-matrix results available. Typical
examples are 4-D gauged AFT with
$n_f$ massless fermions, like QCD, where the  expected\cite{DCSB} $SU(n_f)_L
\times SU(n_f)_R \to SU(n_f)_V$ breaking is characterized via non-perturbative
order parameters, generalizing the role of the mass gap in simpler models
of dynamical symmetry breaking\cite{NJL}.  \\

The paper is organized as follows. 
In section 2, we recall the main steps of our construction. We define
the new  perturbative expansion,
keeping as much as possible the discussion general for any AFT but with
specific illustrations in the $O(N)$ GN model. We also give additional 
formulas and some properties which had not been discussed in
refs.~\cite{gn2}--\cite{qcd2} or \cite{KRlet}. 
In section 3,
we briefly recall the usual problems of infrared renormalon singularities,
the resulting perturbative ambiguities, and how these ambiguities are
usually removed by non-perturbative contributions, whenever the latter can be
explicitly evaluated, e.g. at the next-to-leading $1/N$ order in 
2-D integrable models like the GN. 
In section 4, we reexamine the Borel convergence properties\cite{KRlet} 
obtained for $F<0$ in the simplest situation, providing also a
more detailed analysis of some technical issues. Section 5 and 6 analyse how
the  variationally improved perturbation (DE-VIP) can further improve these
Borel convergence properties, and in section 6 are also introduced other
(non-linear) convenient generalizations of the simpler DE-VIP construction,
which can lead to a directly convergent alternative series. Finally
in section 7 we give numerical applications for the $O(N)$ GN model,
where we compare in some details different 
approximations and/or resummation methods 
which can be constructed order by order within our approach.
Section 8 contains some conclusions, and a number
of technical issues used in various parts of the paper are discussed  
in five appendices.
%%%%%%%%%%%%%%%%%%%%%%%%%%%%%%%% 
\section{RG self-consistent mass and alternative expansion} 
\setcounter{equation}{0} 
 
We consider from now the massive $O(2N)$ 
Gross-Neveu (GN) model\cite{GN,gnnext}, though most of the
discussion and equations are kept general, applying a priori
with minor adaptations to other AFT models,
as long as the low orders RG properties are know. 
The GN
Lagrangian reads 
\bea
{\cal L}_{GN} = \overline{\Psi} i \partial \hspace{-0.2cm} / 
\Psi - m \overline{\Psi}\Psi + \frac{g^2}{2} (\overline{\Psi}\Psi)^2
\label{LagGN}
\eea
where $g^2$ is the four-fermion
coupling  and $m$ an explicit fermion mass. In the $D=2$ GN
$O(2N)$ model, the discrete chiral symmetry $\Psi \to \gamma_5 \Psi$
of the Lagrangian in the $m\to 0$ 
limit, is spontaneously broken\cite{GN} for any $g^2>0$.
The model develops a non-trivial mass-gap $M^P(N)$, whose 
exact expression in the massless limit $m \to 0$ has been established for 
arbitrary $N$ by using the exact S-matrix 
results\cite{exactS} and thermodynamic Bethe Ansatz methods\cite{FNW}. 
In this paper we shall mainly (but not only) work in the so-called 't Hooft
renormalization scheme\cite{thooft}, where  
\be
\beta(g^2)\equiv dg^2/d\ln \mu =
-2b_0 g^4 -2b_1 g^4\;,
\label{betuni}
\ee
(also defining our RG coefficient conventions, so that $b_0 >0$ for an AFT).
This is motivated by the fact that all higher order coefficients 
$b_i$ for $i \ge 2$ are non
universal, being explicitly renormalization scheme (RS) dependent.
Similarly we truncate the anomalous mass dimension to two-loop order:
\be
\gamma_m(g^2) \equiv -d(\ln m)/d\ln\mu = \gamma_0 g^2 +\gamma_1 g^4\;.
\label{gamuni}
\ee
 In the $O(2N)$ GN
model, the RG coefficients $b_i$ and $\gamma_i$ are exactly known in the
$\overline{MS}$ scheme
up to three loop order~\cite{Gracey3loop}: 
\beq
\label{bi}
b_0 = \frac{2N-2}{4\pi},\;b_1 = -\frac{2N-2}{8\pi^2}, \;
b^{\MS}_2 = -\frac{(2N-2)(2N-7)}{64\pi^3} ;
\eeq
and
\beq
\label{gi}
\gamma_0 = \frac{2N-1}{2\pi},\; \gamma^{\MS}_1 = -\frac{2N-1}{8\pi^2}, \;
\gamma^{\MS}_2 = -\frac{(2N-1)(4N-3)}{32\pi^3} .
\eeq
where $\gamma_0$ is universal while, as indicated, the 
$\gamma_1$ coefficient is RS-dependent, as discussed
in more details later (see also Appendix
\ref{ApN2}). 
\subsection{Pole mass and $1/F$ expansion properties}
In the above scheme (\ref{betuni}),(\ref{gamuni}),
one can write the GN pole
mass at arbitrary perturbative orders in terms of the scale invariant mass
$\hat m$, as follows\cite{gn2}--\cite{qcd2},\cite{KRlet}: 
\bea 
 M^P (\hat m) 
 = 2^{-C}\:\hat m  
F^{-A} [C + F]^{-B}   
\; \left[1+\sum^\infty_{n=1} d_n\:(2 b_0 F)^{-n}\;\right], 
\label{MRGn} 
\eea    
with $\hat m$ the scale invariant mass at
second RG order:
\be
\hat m \equiv m(\mu) (2b_0 g^2(\mu))^{-A} [1+\frac{b_1}{b_0}g^2]^B\;,
\label{m2def}
\ee
\beq 
F(\frac{\hat m}{\Lam}) \equiv \ln [\frac{\hat m}{\Lam}] -A  \ln F -(B-C) \ln [C
+F],  \label{F2def} 
\eeq 
related to the usual perturbative coupling as $F^{-1} = 2b_0 g^2(M_{RG})$
where $M_{RG}$ is the pure RG
resummed mass:
\be
M_{RG}(\hat m)  \equiv 2^{-C}\hat m  
F^{-A} [C + F]^{-B}\;.  
\label{MRG}
\ee 
$\Lam$ is the basic scale at second RG order in the 
$\MS$ scheme:
\be
\ds \Lam = \mu\;e^{-\frac{1}{2b_0g^2(\mu)}}\; 
(b_0\,g^2(\mu))^{-C}\:[1+\frac{b_1}{b_0}g^2]^C \;,
\label{Lamdef}
\ee
and
the parameters $A,B,C$ in Eqs. (\ref{MRGn})--(\ref{Lamdef}) are given in terms
of the one-loop and two-loop RG coefficients:
\be
A =\frac{\gamma_1}{2 b_1},\:B =\frac{\gamma_0}{2 b_0}-A,\; C =
\frac{b_1}{2b^2_0}\;, 
\label{ABCdef}
\ee
while the coefficients $d_n$ in Eq.~(\ref{MRGn}) are essentially made of 
the non-RG (non-logarithmic), purely perturbative contributions from the
$n$-loop graphs (generically dominant, as will be discussed
later), plus eventually (subdominant) contributions from higher
RG orders in a scheme with $b_i, \gamma_i \neq 0$ for $i \ge 2$. In the $O(2N$)
GN model, those perturbative coefficients are only known exactly to two-loop
order, and indeed similarly in most of the models. One finds\cite{gn2,AcoVer}
in the  $\MS$ scheme\footnote{We quote in Eq. (\ref{d1d2GN}) the exact value
of $d_2$, recently obtained in the second ref. of \cite{AcoVer}. The tiny
difference with the former (numerical) value as given in \cite{gn2}: $d_2 \sim
(0.737775 -\pi^2/96)\,(2N-1)/(2\pi^2)$, does not affect any of our numerical
results.}:  
\be
\ds
d_1 = 0\;;\;\;\;d_2 = (\frac{\zeta[2]}{2} -\frac{3}{16})\:
\frac{(2N-1)}{2\pi^2}\;.
\label{d1d2GN}
\ee
Eqs.~(\ref{MRGn})--(\ref{Lamdef})
 were obtained by integrating exactly the usual 
RG evolution for the (renormalized) 
Lagrangian fermion  
``current" mass:  
\beq 
m(\tilde \mu) = m(\mu )\;\; {exp\{ -\int^{g(\tilde \mu)}_{g(\mu )} 
dg {\gamma_m(g) \over {\beta(g)}} \} }\;, 
\label{runmass} 
\eeq 
using the   
{\em self-consistent} condition
$M_{RG} \equiv m(\tilde \mu \equiv M_{RG}) $
defining $M_{RG}$ in Eq.~(\ref{MRG}). 
Note that this resummation of the pure RG dependence, exact at
second RG order (and to all orders in the two-loop truncated 
scheme (\ref{betuni}), (\ref{gamuni}), 
 is explicitly factored out in
Eq.~(\ref{MRGn}) from the  purely perturbative series $\sum d_n/(2b_0 F)^n$:
the resummation of the latter series is precisely the non-trivial
issue in most renormalizable models, as discussed in details later.\\ 
One can easily check that  
(\ref{MRGn})--(\ref{MRG})  
are scale invariant expressions, 
by construction to 
all orders\footnote{Again ``all orders"  
at this stage means within the scheme (\ref{betuni}), (\ref{gamuni}).}.
Eq.~(\ref{MRGn}) is perturbatively consistent, for $\hat m \gg \Lam$, with the
usual expansion relating the current mass at the scale $M^{pole}$ and  
$M^{pole}$ itself\cite{Tarrach,Broadhurst}: 
\beq 
M^{pole} = m(M^{pole}) [1 +\sum_{n=1}^{\infty} c_n g^{2n}(M^{pole}) ] 
\label{mpolegen} 
\eeq 
where the $c_n$ coefficient are related to the $d_n$ ones in Eq.~(\ref{MRGn})
in an easily calculable way, involving also RG-dependent quantities, whose
precise expressions are unessential here.\\

\begin{figure}[htb]
\begin{center}
%\vspace{-2cm}
\mbox{ 
\psfig{figure=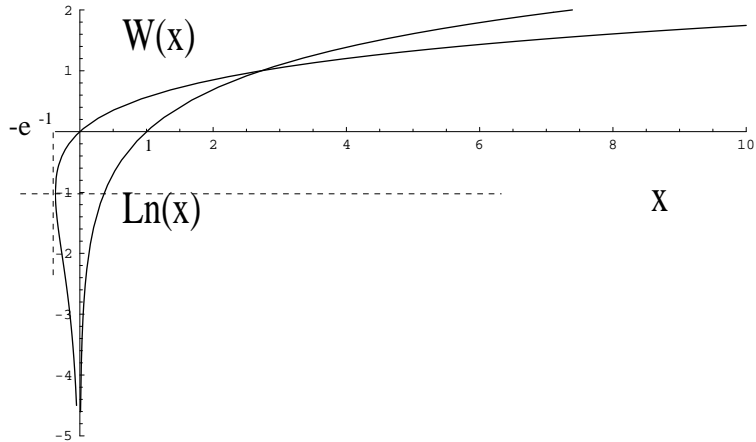,width=10cm}}
\end{center}
\vspace{-3cm}
\caption[]{\label{lambert} The Lambert W function compared to the Log.}
\end{figure}

We concentrate now for the time being on the properties of the pure
RG-dependent, resummed mass expression, Eq.~(\ref{MRG}). At first RG order
($b_1=\gamma_1=0$) Eq.~(\ref{F2def}) takes the simpler form
\be
F(\hat m/\Lam) \equiv
\ln (\hat m/\Lam) -A_0 \ln F
\;= A_0\: W[A_0^{-1} 
(\hat m/\Lam)^{1/A_0} ]
\label{Fdef}
\ee 
where the Lambert\cite{Lambert} function $W[x] \equiv \ln x - \ln W$,
is plotted in Fig~\ref{lambert}, 
and
\be
A_0 = \frac{\gamma_0}{2b_0}\;\; = \frac{2N-1}{2N-2}\;\;
\raisebox{-0.4cm}{~\shortstack{ $\to$ \\ $N\to\infty$}} 1
\label{A0def}
\ee
(where the specific value of $A_0$ for the  $O(2N)$ GN model is
indicated for illustration).  
%%%%%%%%%%%%%%%%%
%  FIGURE 1
%%%%%%%%%%
Eq.~(\ref{Fdef}) has the remarkable property: 
\be
F 
\raisebox{-0.4cm}{~\shortstack{ $\simeq$ \\ $\hat m\to 0$}}
(\hat m/\Lam)^{1/A_0}
\label{Flim0}
\ee
for 
$\hat m \to 0$, in contrast with the ordinary logarithm function (see Fig.~1),
but the latter is asymptotic to $F(\hat m/\Lam)$ for $\hat m \gg \Lam$.
More precisely, on its principal branch (defined to be the one real-valued
for real $\hat m$), $F$ has an alternative
power series expansion\cite{Lambert,KRlet}:
\be
F(x) = \sum^\infty_{p=0} (\frac{-1}{A_0})^p 
\frac{(p+1)^p}{(p+1)!}\; x^{\frac{p+1}{A_0}}
\label{Fexp}
\ee 
which has a finite convergence radius $R_c = e^{-A_0} (A_0)^{A_0}$,
of order ${\cal O}(1)$:
for example in the $O(2N)$ GN model, $R_c \sim 0.41$ 
[$R_c = e^{-1}\sim 0.37$] for
$N=2\;[N \to\infty]$.
%%%%% FIGURE  2 %%%%
%\vspace{-2cm}
\begin{figure}[htb]
\begin{center}
\mbox{
\psfig{figure=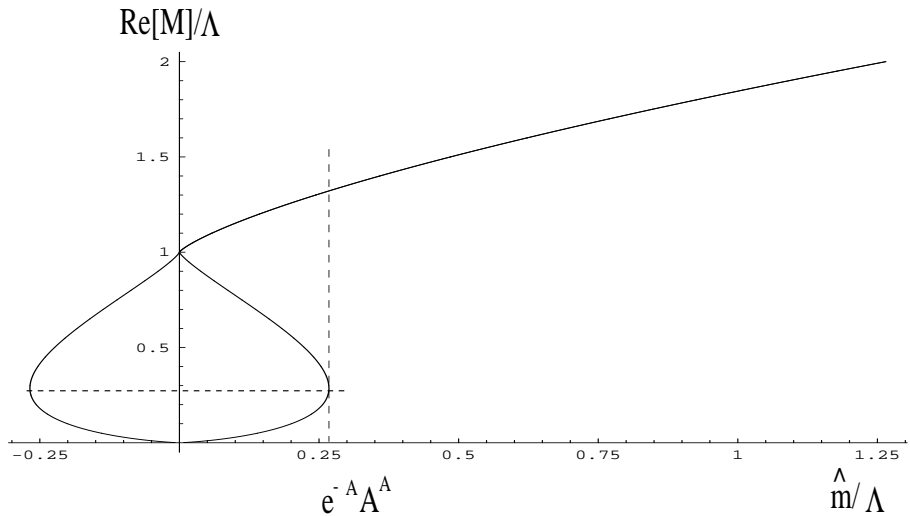,width=12cm,height=12cm}}
\vspace{-3cm}
\end{center}
\caption[]{\label{mongolfier} The different branches of 
$Re[M(\hat m)]/\Lam$ at first RG order in Eq.~(\ref{MRG}), 
determined by the value of $A \equiv A_0$ in Eq.~(\ref{A0def}), for 
$N=3$ in $O(2N)$ GN model.
Similar branch structures occur at second RG order, for
Eq.~(\ref{MRG}) with $b_1, \gamma_1 \neq 0$, and as well in other AFT.}   
\end{figure}
%%%%%%%%%%%%%%%%%%%%%%%%%%%
The pure RG mass $M_{RG}(\hat m)$ in Eq.~(\ref{MRG})   
thus exhibits different branches,
determined both by the values of the RG parameter 
$(-1)^{A_0}$ and by the original two branches of the
Lambert function (see Figs.~\ref{lambert} and
\ref{mongolfier}). 
In Fig.~\ref{mongolfier} the two principal branches, extending
for $Re[M]/\Lam <1$, correspond to $F <0$  with $\hat m >0$
or $\hat m <0$ respectively, as given by the roots of
$(-1)^{A_0} = (-1)^{5/4}$ e.g. for $N=3$
in the $O(2N)$ GN model. Each of these branches divides 
in turn into two
branches, for  $e^{-A_0}\,\Lam <Re[M]<\Lam$ and $0<Re[M]<e^{-A_0}\,\Lam $, 
respectively, which correspond to the original branch structure of the
Lambert function, above and below $W = F=-1$, see Fig.~\ref{lambert}. Clearly,
the physical branch is the one with $\hat m >0$, $\Lam <Re[M] <+\infty$. It is
indeed the only one branch which for {\em real} $\hat m$ values, is real and
continuously matching the asymptotic perturbative 
behaviour of $F$ at large $\hat m$. Moreover it is the branch 
consistent with a 
non-zero ``mass gap" $M_{RG} = \Lam$ for $\hat m \to 0$. 
Algebraically, this RG mass is
obtained by expanding Eq.~(\ref{Fexp}) to first orders in
(\ref{MRG}):   
\be
M_{RG}(\hat m \to 0) = \hat m \;[(\hat m/\Lam)^{1/A_0}+\cdots]^{-A_0}\;
=\: \Lam \;(1+{\cal O}(\hat m/\Lam)^{1/A_0})\;,
\label{M1Lam}
\ee
which can be viewed as a generalization 
(for $m \neq 0$) of dimensional transmutation\cite{ColWei}: more precisely,
using Eq.~(\ref{Fexp}) one can easily get systematical corrections in 
powers of $(\hat m/\Lam)^{1/A_0}$ to the $M_{RG}/\Lam$ ratio in
Eq.~(\ref{M1Lam}). Eq.~(\ref{M1Lam}) automatically
reproduces, e.g., the GN $O(N)$ model mass gap in the large $N$,
$m\to 0$ limit (where $A\to 1$ for $N\to\infty$),
traditionally obtained in a different way~\cite{GN}. 
The multivaluedness of $F$ and $Re[M]$ in
the infrared region does not signal ambiguities for the physical quantities:
actually the different branch structure is completely fixed for given RG
coefficients.  This is generic, so that
a similar structure occurs e.g. in QCD, but with of course a different value
of the RG parameter $A_0$\cite{KRlet}. We again stress, however, 
that Eq.~(\ref{M1Lam}) alone is not a proof of dynamical
\CSB, as one may naively infer from the previous discussion: in fact, 
our construction only exploits
that (from dimensional
transmutation) any mass is
proportional to $\Lam$ for $\hat m \to 0$ in an AFT, 
which is encoded here in
the properties of  $F(\hat m)$ for any $\hat m$. In particular we emphasize 
that the same properties still hold e.g. for any 2-D AFT models
with a continuous chiral symmetry, where spontaneous breakdown
is not possible according to Coleman theorem\cite{Colth}, but where a 
physical mass gap occurs\cite{TBArev}. \\  
At second RG order, $F$ as defined by Eq.~(\ref{F2def}) cannot be written in
terms of the Lambert function, but it has very similar properties which
can be  easily inferred from its reciprocal function:
\be
\frac{\hat m}{\Lam} = e^F F^A (C+F)^{B-C}.
\label{recipro}
\ee  
Indeed, replacing Eq.~(\ref{recipro}) in Eq.~(\ref{MRG}) immediately gives
a simpler expression for the pure RG mass, as a function of $F$:
\be
M_{RG}(F) =2^{-C}\:e^F\:(C+F)^{-C} \; \Lam 
\label{MRGF}
\ee 
which thus only depends on the universal RG quantity $C$ defined in
Eq.~(\ref{ABCdef}). For illustration
we plot in Fig~\ref{Fplot} the function $F$ defined in
Eq.~(\ref{F2def}), for the $O(2N)$ GN model case with $N=2$, where the
relevant RG coefficients $b_0$, $b_1$, $\gamma_0$, $\gamma_1$ were
defined in Eqs~(\ref{betuni}), (\ref{gamuni}). 
%%%%% FIGURE  3 %%%%
\begin{figure}[htb]
\begin{center}
\mbox{
\psfig{figure=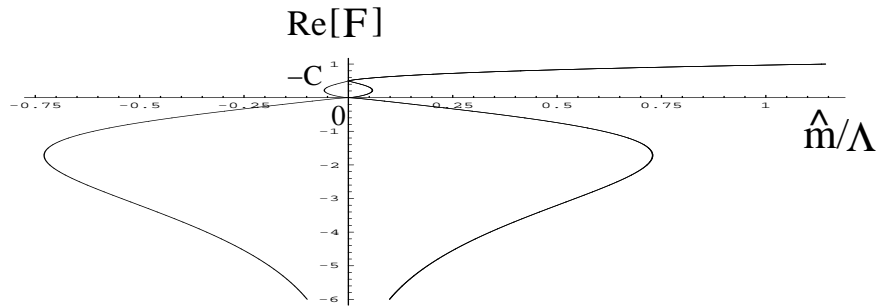,width=12cm}}
\vspace{-2cm}
\end{center}
\caption[]{\label{Fplot} The different branches of 
$F$ at second RG order in Eq.~(\ref{F2def}) in the $O(2N)$ GN ($N=2$).}   
\end{figure}
%%%%%%%%%%%%%%%%%%%%%%%%%%%
Similarly to first RG order in Eq.~(\ref{Fexp}), $F$ has a power
expansion for sufficiently small $\hat m$: defining
\be
F(x) \equiv A\: G[\: \frac{C^{(B-C)/A}}{A}\: x^{1/A}] 
\label{FRG2}
\ee 
with $x =\hat m/\Lam$, one has from (\ref{F2def}) 
\be
G = x\; e^{-G} (1+\frac{A}{C}\:G)^{C-B} \equiv x \;
[1+\sum^\infty_{p=1} a_p x^p ]
\label{G2d}
\ee
where the expansion coefficients $a_p$ are now more involved 
than the first order corresponding ones in (\ref{Fexp}) (the $a_p$ 
will now depend explicitly on the RG
quantities $A$, $B$, $C$), but can be derived
systematically~\footnote{The
$a_p$ in Eq.~(\ref{G2d}) are easily evaluated to high order with e.g.
Mathematica\cite{matha}.}. 
It is easy to determine the convergence
radius of this expansion form of $F(\hat m)$ around zero, given by the location
of  its closest non-zero singularity\footnote{In Eq.~(\ref{F2sing}) 
the solution $ 1 - \sqrt{\cdots}$ is only valid for $\gamma_1 \neq 0$:
if $\gamma_1 = 0$, $F^* =0$ is not a singularity, as is clear from the first RG
order result, where the only singularities ly on the circle of radius
$e^{-A_0} (A_0)^{A_0}$.}:  
\be
\frac{d\,\hat m}{dF}(F^*) =0 \to F^* = -\frac{1}{2}\;(\frac{\gamma_0}{2b_0})\;
[1 \pm \sqrt{1-4\frac{\gamma_1}{\gamma^2_0}}\;\;]\;,
\label{F2sing}
\ee
and correspondingly     
\be
\frac{\hat m^*}{\Lam} = e^{F^*} (F^*)^A (C+F^*)^{(B-C)}
\label{Rc2}
\ee
so that $R_c \equiv |m^*/\Lam|$. Note that for $N\to\infty$,
$F^*=-1$ and correspondingly $\hat m^*/\Lam =-e^{-1}$. This is due to
the fact that for $N\to\infty$ $F$ is exactly the Lambert function, see 
Fig.~\ref{lambert}. At second RG order, in
the $O(2N)$ GN model, $F^*<0$ and $C+F^*<0$ are real and negative 
$\forall N
\ge 3/2$, so that the number and location of the singularities in
Eq.~(\ref{Rc2}) on the circle of radius $R_c$ are determined by the roots of
$(-1)^{A+B-C}$, see  Fig.~\ref{convrad}.\\  
%%%%%%%%%%%%%%%%%
%  FIGURE 4
\begin{figure}[htb]
\begin{center}
\vspace{-4cm}
\mbox{ 
\psfig{figure=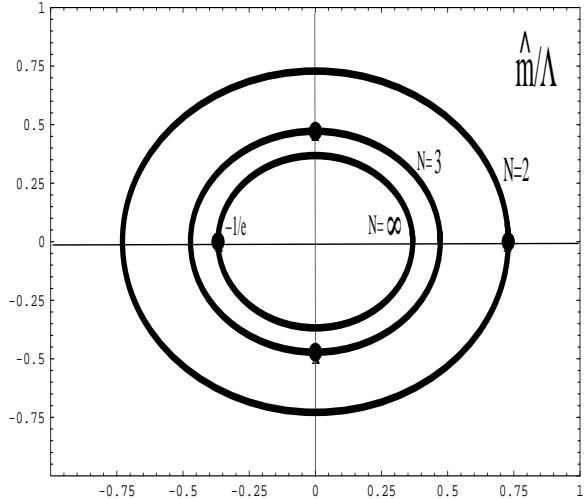,width=8cm,height=16cm}}
\end{center}
\vspace{-5cm}
\caption[]{\label{convrad} Convergence radii of the
power expansion form Eq.~(\ref{G2d}) of $F$ at second RG order,
Eq.~(\ref{F2def}), and 
corresponding 
singularity locations on the convergence circle for different values of 
$N$ in the $O(2N)$ GN model.}
\end{figure} %%%%%%%%%%
\subsection{Other renormalization schemes}   
Before to proceed we should remark that at second RG order there is 
a certain arbitrariness in e.g. Eq.~(\ref{F2def}) and related quantities,
like (\ref{F2sing})--(\ref{Rc2}), since  
$\gamma_1$ is renormalization scheme (RS)-{\em dependent}:
more precisely, for an arbitrary perturbative RS change in the
Lagrangian mass and coupling parameters:
\bea 
 g^2 \to \tilde g^2 = g^2\;(1 +A_1 g^2  +\cdots)\; , \nn \\
 m \to \tilde m = m \;(1 + B_1 g^2 + \cdots) \; , 
\label{RSchange}
\eea
$\gamma_1$ is changed as
\be
 \tilde \gamma_1  = \gamma_1 +2b_0 B_1 -\gamma_0 A_1\; 
\equiv \; \gamma_1 +\delta\gamma_1\;, 
\label{RSC}
\ee
while $b_0$, $b_1$ and $\gamma_0$ 
are RS-invariant, as already mentioned 
(more details on RS changes are given in Appendix C).
This means for instance that the location of the 
singularities and the value of the convergence radius as implied by
(\ref{Rc2}) may be modified, to some extent, by appropriate 
changes in $\gamma_1$ (equivalently changes in the quantity $A \equiv
\gamma_1/(2b_1)$ as defined in (\ref{ABCdef})).\\ %
It is always possible to choose the `t Hooft scheme, in which
Eqs.~(\ref{MRGn}), (\ref{MRG}), (\ref{F2def}) resum 
the complete RG dependence in $\hat m$. This is convenient because 
beyond second RG order, in an arbitrary scheme where $b_i, \gamma_i \neq 0$
for $ i \ge 2$,
algebra becomes quite involved, and neither the non-log contributions $d_n$
nor the $b_n$ and $\gamma_n$ RG coefficients are known at arbitrary
orders for most field theories, and in particular for the GN model. 
Nevertheless it is still possible to work out a
formal generalization
of Eq. (\ref{MRGn}), in an arbitrary (MS) scheme, see Appendix A. 
We will use
this generalization to define some of the numerical approximations to the mass
gap,  of arbitrary higher orders, in section \ref{Num}.\\ 
Before to conclude
this section, we discuss another possible RS choice, obtained
from expression (\ref{MRGn}) by an all orders redefinition
of $F$:
\be
F \equiv \tilde F -(A+B-C) \ln \frac{F}{\tilde F} -(B-C) \ln [1+\frac{C}{F}]
\label{NPRS}
\ee
which can be perturbatively expanded in powers of $1/\tilde F$, where
$\tilde F$ is now again directly related to the Lambert function: 
\beq 
\tilde F(\frac{\hat m}{\Lam}) \equiv \ln [\frac{\hat m}{\Lam}] -(A+B-C)  \ln
\tilde F\; \equiv \tilde A W[(\frac{\hat m}{\Lam})^{1/\tilde A}/\tilde A]
\label{FGNsc} 
\eeq  
with $\tilde A\equiv A+B-C$.
This redefinition is motivated from the fact that
in the GN model, the RG coefficient $C <0$ in
Eq.(\ref{ABCdef}),
due to $b^{GN}_1 <0$, which corresponds to an infrared fixed point
at $g_c = -b_0/b_1 >0$, so that the perturbative branch of $F$ reaches
$\hat m \to 0$ first for $F=-C$. In the scheme (\ref{NPRS}), Eq.~(\ref{MRGn})
takes the form:
\bea 
 M^P (\hat m) 
 = (2/e)^{-C}\:\hat m \; 
\tilde F^{-(A+B)}   
\; \sum^\infty_{n=0} \tilde d_n\:(2 b_0 \tilde F)^{-n}\;. 
\label{MRGNsc} 
\eea    
This also implies appropriate changes in the purely perturbative coefficients, 
simply determined
by re-expanding Eq.~(\ref{F2def}) in $1/\tilde F$ powers:
\be
\tilde d_1 = d_1 - B\;C\;,\;\;\cdots
\label{dnGN}
\ee
\section{Infrared renormalon properties of the GN pole mass}
\setcounter{equation}{0}
As mentioned in introduction, the idea is that, since the complete pole
mass Eq.~(\ref{MRGn}) gives the ratio $M^P(F)/\Lam$ to all perturbative
orders for $\hat m \gg \Lambda$, if we are able to resum this series {\em and}
to give it a meaning for $\hat m \to 0$, we can obtain the
$M^P/\Lam$ ratio in the physically interesting massless limit. As far as the
pure RG dependence is concerned, this turns out to be possible because $F(\hat
m)$ provides a rigorously defined and explicit bridge between the
``non-perturbative" $\hat m \lsim \Lam$ regime,  where $F$ has power
expansion (\ref{Fexp}),
and the short distance perturbative $\hat m \gg \Lam$ (logarithmic) regime. 
A crucial point indeed is the difference between the usual effective
coupling $g^2(p^2) \equiv 1/[b_0 \:\ln(p^2/\Lam^2)]$, having a Landau pole
at $p^2 = \Lam^2$, and $F^{-1}(\hat m)$ here, 
having its pole at
$\hat m =0$, governing the massless limit (\ref{M1Lam}) of the 
(pure RG) mass gap Eq.~(\ref{MRG}).
Accordingly along the continuous branch on Fig.~2, $M(\hat m)$ has
no singularity for $0 <\hat m <\infty$, as is clear
also from Eq.~(\ref{M1Lam}) and Fig.~1, 2. 
Now, to extrapolate
the complete pole mass (\ref{MRGn}) down to the chiral, strongly coupled regime
$\hat m \simeq 0$, the main obstacle comes from the presence of the 
purely perturbative coefficients $d_n$. First, though
the pole mass (or other physical quantities similarly)
 is infrared finite, gauge~\cite{Tarrach}--,  scale-- and scheme--invariant, 
the relation between the pole mass and e.g. the running mass in
(\ref{mpolegen}) is scheme dependent, which is manifested here by the
RS-dependence in (\ref{MRGn})   of the perturbative coefficients $d_n$,
the RG coefficients $A$, $B$
in Eq.~(\ref{ABCdef}),  
and of $\Lam$ too.\\
Second, it is immediate that the perturbative
contributions $d_n/F^n$ in Eq.~(\ref{MRGn}) are singular
when $F \to 0$ ($\hat m \to 0$), since each term will
have a leading divergence $\simeq d_n \;(\hat m/\Lam)^{-n/A}$, according
to (\ref{Fexp}). In other words, while the usual Landau
pole problem was avoided
in the pure RG part (\ref{MRG}) of $M^P$, which has a regular finite $\hat m
\to 0$ limit, as illustrated in Figs.~1--3, a problem reappears in the
perturbative corrections relating the true physical quantities, like the pole
mass, to their pure RG part. Thus, in an arbitrary scheme, 
strictly speaking $M^P \to \infty$ when $\hat m \to 0$.  
(In principle one could avoid this problem in a crude way by exploiting the RS
arbitrariness in (\ref{MRGn})
to define a scheme such that {\em all}
the perturbative coefficients $\tilde d_n \equiv 0$. Although such a 
peculiar scheme can always be formally constructed, this 
solution is to be considered unsatisfactory, since one expects
truly non-perturbative results not to depend on a particular
scheme.) This appears in fact  
completely similar to the perturbative expansion in powers of
$(g/m^3)$ of the oscillator energy levels, thus also singular for 
$m\to 0$, which nevertheless do not
prevent different resummation methods to work very
well\cite{ao},\cite{delta}--\cite{deltac}, even for $m\to 0$. We will see
in next sections how similar resummation properties generalize, to
some extent, in the present field theory case.\\
%%%%%%%%%%%%%%%%%
%  FIGURE 4
\begin{figure}[htb]
%\begin{center}
\vspace{2cm}
\hspace{-6cm}
\mbox{ 
\psfig{figure=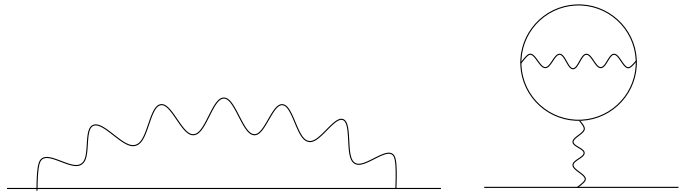,width=2truecm}}
%\end{center}
%\vspace{-2cm}
\caption[]{\label{GNmass} The GN mass graphs at order $1/N$.}
\end{figure}
%%%%%%%%%%
Now there is unfortunately an even worse problem, when dealing with
the purely perturbative expansion of the pole mass: in the $O(N)$ GN model,
at order $1/N$, it exhibits infrared 
renormalons very similar to the QCD quark pole
mass\cite{MPren,renormalons}. More precisely,
let us consider only the naive perturbative
expansion of the pole mass, obtained from Fig.~\ref{GNmass}
by taking perturbative expression of the dressed scalar propagator (wavy line),
$G^{-1}(q^2) \sim g^2(q^2) \sim [ b_0\:\ln (q^2/\Lam^2)]^{-1}$:
\be
M^P = 1+\frac{1}{4N} \int^{\mu^2}_0 \frac{d q^2}{M^2} (1-\zeta)\; 
\left[g^2(q^2) -{\cal O}(\frac{M^2}{q^2}) \right]
\label{MP1Npert}
\ee
where
$\zeta = (1+4M^2/q^2)^{1/2}$ and $M \equiv \mu e^{-1/(2b_0\:g^2(\mu))}$
is the mass gap at leading $1/N$ order.
Then a standard calculation gives\cite{KRcancel}
\be
d_{n+1}
\raisebox{-0.4cm}{~\shortstack{ $\sim$ \\ $n\to\infty$}}
(2b_0)^n\: n!
\label{irren} 
\ee  
so that the series Eq.~(\ref{MRGn}) including this next-to-leading
$1/N$ order is badly
divergent for any $\hat m$, and not
even Borel summable: such a factorial growth of the
perturbative coefficients, with no sign alternation,
implies\cite{renormalons}
ambiguities of ${\cal
O}(\Lam)$. 
But, those renormalons are only perturbative
artifacts: considering now the full scalar propagator
contribution (which is known exactly for the $O(N)$ GN model at $1/N$ order):
\be
G[q^2] = \left[1+ \frac{g^2\, N}{2\pi}\left[\ln\frac{M^2_P}{\mu^2} +\zeta
\ln[\frac{\zeta+1}{\zeta -1}]\:\right] \right]^{-1} 
= \left[\frac{g^2\, N}{2\pi}\zeta \ln[\frac{\zeta+1}{\zeta -1}] \right]^{-1} 
+{\cal O}(1/N)\;,
\label{Gq2}
\ee
rather than its truncated 
perturbative contribution Eq.~(\ref{MP1Npert}), the exact $1/N$ expression of
the pole mass is obtained\cite{KRcancel,FNW} as
\be
M^P = \Lam \left[ 1+\frac{1}{2N} [ Ei[-\theta] -\ln \theta 
-\gamma_E +\ln(\ln \frac{\mu^2}{M^2}) -2 \ln (\cosh[\theta/2])+\ln
\frac{\mu^2}{M^2}\;] \right] 
\label{MexactEi}
\ee
with $\chi=(1+4 M^2/\mu^2)^{1/2} \equiv 1/\tanh(\theta/2)$ (i.e.
$\theta = \ln[(\chi+1)/(\chi-1)]\;\ge 0$), 
and $Ei(-x) \equiv -\int^\infty_x dt e^{-t}/t$ the Exponential
Integral function ($x \ge 0$). Thus $Ei[-\theta]$ has
a factorial perturbative series with {\em sign-alternated} coefficients,
i.e. the IR renormalons actually disappear:
more precisely we can re-expand the result (\ref{MexactEi}) in
perturbation, using 
\be
\ln[\frac{\chi+1}{\chi -1}] \simeq \ln \frac{\mu^2}{M^2} +2 \frac{M^2}{\mu^2} 
+\cdots = g^{-2}+2 \frac{M^2}{\mu^2} +\cdots 
\ee
\be
M^P = M \left[1+ \frac{1}{2N} \left( 2\ln 2 -\gamma_E  
-\frac{M^2}{\mu^2}  [\sum^\infty_{n=0} (-1)^{n} n!\;g^{2(n+1)}\; +{\cal O}
(\frac{M^2}{\mu^2})]
\right) \right] 
\label{genuine}
\ee
The explicit Borel summability of the genuine perturbative expansion,
Eq.(\ref{genuine}), is not in contradiction with the purely perturbative
results above, because the non-trivial cancellation of renormalons involve
the contributions of non-perturbative power corrections
contributions\cite{KRcancel,David,Benbraki}. Moreover, it turns out that 
this cancellation is such that the final expression of the pole mass
contains neither ``purely perturbative" nor ``intrinsically non-perturbative"
contributions: for instance, the net contribution due to the first graph in
Fig. \ref{GNmass} is the term $-\gamma_E$ in Eq.~(\ref{genuine}), which simply
remains after cancellation of the first order terms:
\bea
\ds
& &
-\int_0^\infty dt 
\left[\frac{1}{t(1+t)} 
+e^{-t/g^2}[ -\frac{1}{t} +{\cal O}(\frac{M^2}{\mu^2})\:]\right] \nn \\
&\sim &
-\frac{1}{2} [\gamma_E +{\cal O}(\frac{M^2}{\mu^2})]
\label{MBresult}
\eea
Now, the point is that the above results Eqs.~(\ref{MexactEi}),
(\ref{genuine}) obviously could only be obtained from calculating explicitly 
the exact mass gap at next-to-leading $1/N$ order. Our aim here is to ignore on
purpose these exact $1/N$ results, a priori only accessible in a certain class
of 2-D models. Rather, we want to examine whether our generic construction,
relying solely on the purely perturbative information, together with the
infrared properties of the function $F$, is able to recover some of the
non-perturbative properties of the exact mass gap, in particular its Borel
summable asymptotic expansion explicit at  next-to-leading $1/N$ order. 
\section{Borel summability of the $F$ expansion}
We first reexamine here why the expression 
(\ref{MRGn}) is plagued with perturbative ambiguities, and how one can get rid
of those, within our construction, thanks to the analytic properties of $F$ in
a vicinity of
$\hat m = 0$ values\cite{KRlet}. First we define from Eq.~(\ref{MRGn}) its 
Borel transformed 
series\footnote{The resummed RG-dependence  $\hat m
F^{-A}(C+F)^{-B}$, having obviously no factorial
behaviour, is thus
factored out of the Borel transformed series.}:
\be
B.T.[M^P(F)](t) \equiv  2^{-C}
\hat m\:F^{-A}(C+F)^{-B}\;\left[1+\sum^\infty_{n=1} \frac{d_n\:t^n}{(n-1)!}
\right]
\label{BT}
\ee
so that the corresponding Borel integral reads: 
\be
BI(\hat m) \equiv \tilde M^P(\hat m)   =  2^{-C}
\hat m\:F^{-A}(C+F)^{-B}\int^\infty_0 \! dt e^{-t}\;
[1 +(4\pi\,b_0\,F)^{-1}\sum^\infty_{n=0} (\frac{t}{F})^n \:] 
\label{BI}
\ee
upon assuming for the perturbative coefficients $d_n$ 
the leading large order behaviour in Eq.~(\ref{irren}). For any $F >0$, this
expression would be (asymptotically) equal to (\ref{MRGn}) by formal
expansion,  would the pole at  $t=F$ not make the integral (\ref{BI})
ill-defined. One should make a choice in e.g. deforming the contour above
(or below)  the pole,
which results in an ambiguity, which is easily seen to be proportional 
to ${\cal
O}(e^{-F})$. Since from Eq.~(\ref{Fdef}) $F \sim \ln [\hat
m/\Lam]-A \ln [\ln [\hat
m/\Lam]]$ for $\hat m \gg \Lam$, this implies a perturbative ${\cal
O}(\Lam/m)$  ambiguity for the ``short distance" ($M, \hat m \gg
\Lam$) pole mass:
\be
\mbox{ambig}\;\sim
\pm i\;(\frac{\Lam}{\hat m})\;\ln^A[\frac{\hat m}{\Lam}]\;,
\ee
in consistency with general
results\cite{renormalons}. Now in our case, Eq.~(\ref{Flim0}) (and equivalently
Eqs.(\ref{FRG2}),(\ref{G2d}) at second RG order) allow  to trace the
behaviour of $F$ all the way down to $\hat m\to 0^+$, where at first RG order,
$F\to 0$: consequently the naive mass gap (\ref{MRGn}), expected to be $\sim
\Lam$, is also ambiguous by ${\cal O}(\Lam)$.\\ 
But in contrast, within our construction, the Borel integral
(\ref{BI}) can be defined unambiguously and independently of the RS
parameter $A$, in the range $F <0$\cite{KRlet}: then  $F \equiv -|F|$
simply produces the adequate sign alternation in the factorially
growing coefficients $\sim F^{-n}$.  More precisely,  a straightforward
calculation of Eq.~(\ref{BI}) for $Re[F]<0$ (neglecting for simplicity at
the moment the two-loop RG dependence $C$, irrelevant to asymptotic
properties), gives  
\be
BI(\tilde M^P/\Lam)  \raisebox{-0.4cm}{~\shortstack{ $\sim$ \\ $F < 0$}}
e^{-|F|}+\frac{1}{2 b_0}\: Ei(-|F|)
\label{directBS}
\ee
where we also used Eq.~(\ref{recipro}) to express $M^P/\Lam$ as a function
of $F$ only. 
Indeed, as already mentioned the first RG order function $F(\hat m)$
in (\ref{Fdef}) is well-defined (analytic) for any $A$ values
in a disc of radius $e^{-A} A^A$ around zero (and for $A=1$ the only
singularity is at $F=-1$ i.e. $\hat m/\Lam = -e^{-1}$, cf. Fig.~\ref{lambert}).
Thus, one can choose the branch of $F$ such that $Re[F(\hat m)] <0$,
compatible with the limit $\hat m \to 0$. 
The second RG order $F$ in  Eq.~(\ref{F2def}) has similar properties, 
with finite convergence domain around
$F=0$ and $F=-C$ respectively, see
Fig~\ref{Fplot}~\footnote{Note that at second RG order, from
Eq.~(\ref{recipro}) the point $F=-C$ also corresponds to $\hat m =0$: we
shall come back on this later on for the GN model, where $F=-C >0$ corresponds
to the infrared fixed point at $g^2 =-b_0/b_1 >0$, as already mentioned.}. More
generally, in an arbitrary AFT with arbitrary values of the RG parameters $A$,
$B$, $C$ depending on the renormalization scheme, there always exist branches
of $F$ such that $F$ is complex. This is the case for the GN model for the two
branches shown with  $Re[F] < -C$ in Fig. \ref{Fplot}. As a consequence, the
Borel singularities in e.g. Eq.~(\ref{BI}) are moved away from the real
semi-axis of  Borel integration $Re[t]>0$, now being located at $t_0 = F
\equiv |F] e^{i\theta_F}$ with $\theta_F  \neq 0$. \\

We obtain in this way formal Borel convergence for a
certain range of the expansion parameter near the relevant massless limit
$\hat m =0$,
strictly only along those branches such that $F<0$, or more generally complex.
Depending on the branch of $F$, this may correspond either to $\hat
m >0$ or $\hat m <0$ (see Figs.~\ref{mongolfier},\ref{Fplot}), which is
in principle not a problem, since relativistic field theories only depend on
$m^2$ (the sign of the Lagrangian mass term in (\ref{LagGN}) can be flipped by
a discrete $\gamma_5$ transformation, and the Dirac equation is invariant
under $m \to -m$). In Fig.~\ref{mongolfier}, the complex branches of $F$ are
those  corresponding to $Re[M]/\Lam <1$, and to $Re[F]<-C$ in Fig.
\ref{Fplot}, where in both cases the
symmetry with respect to $\hat m \to -\hat
m$ of such pure RG dependence is manifest. More generally,
we expect that our final, physical mass gap result, should be
independent of the way in which the massless limit is to be reached, either
from $Re[F]>0$, or $Re[F]<0$, or more generally from any of the complex
branches of $F$.  However, the (unphysical) perturbative expansion is clearly
not invariant under this, since the (usual) expansion with real $F>0$ is non
Borel summable and ambiguous. Therefore, the following picture emerges: in our
construction, the perturbative expansion near the strongly coupled, massless
limit can exist in two modes:\\ 
i) in the standard mode, corresponding to
real $F>0 $, and matching the usual perturbative expansion for
$\hat m \gg \Lam$, the
perturbative expansion alone has to be  necessarily completed
as usual with ``non-perturbative" power corrections, as illustrated 
explicitly with the exact $1/N$ calculation of the GN mass gap,
discussed in section 3. \\
ii) In the ``alternative" expansion mode,
with $F<0$ in the simplest case (first RG order) where $A_0=1$, the
perturbative series is {\em directly} Borel summable, thus non-ambiguous.
There is, therefore, no explicit non-perturbative power correction
contributions needed in principle. Those
results are completelly general for a renormalizable AFT of dimension 
$D \ge 2$, since
they only depend on the generic RG properties of the function $F$. \\
To illustrate perhaps better the last points, we
can give an analogy, to some extent, in the simplest possible model where
such issues can be discussed, the anharmonic oscillator. The
detailed analogy is discussed in Appendix \ref{ao}. The oscillator is
described\cite{ao} by a $g \phi^4$ massive scalar field theory in 1-D,
with 
energy levels having from purely dimensional considerations a perturbative
expansion in powers of $g/|m|^3$:
\be
E_0 \sim |m|\;\sum_n a_n (\frac{g}{|m|^3})^n\;,
\label{E0exp1}
\ee
with factorially growing but sign-alternated
coefficients\cite{ao,zinn}: $a_n \sim (-1)^n n!$ at large orders, thus
the energy levels are Borel-summable\cite{aoBsum}.   
Now, let us assume, momentarily, that our only knowledge of the
oscillator would consist of the perturbative expansion Eq.~(\ref{E0exp1}),
and consider formally changing the sign of the coupling $g$ there:
this obviously induces a change
of sign in the perturbative coefficients, rendering the  corresponding series
non Borel summable~\footnote{This is not to be confused with the double-well
potential, obtained from the oscillator by the change: $m^2 \to -m^2$, which
also has a non Borel-summable perturbation series.}. This accordingly produces
an ambiguity, an imaginary part in the energy, whose leading terms can be
evaluated exactly still using the Borel integral, and according to the standard
interpretation it calls for additional non-perturbative corrections. The latter
are easily shown to have the form of the standard instanton contributions to
the ground-state energy\cite{zinn}: 
\be Im \,E_0 \sim \frac{4}{\sqrt{2\pi}}\;
e^{\frac{4m^3}{3g}}\;(\frac{m^3}{-g})^{1/2} \label{aoinst1} \ee
where $g <0$, and governing accordingly the decay of the wave
function due to barrier penetration~\cite{zinn} 
(see appendix \ref{ao} for more details). 
Of course, this instanton contribution was originally not derived from the
perturbative  ambiguity, since in the oscillator case a ``direct" non
perturbative calculation is possible. But as seen from the ``perturbative
only" side, the above argument indicates that, already for the simpler
oscillator
case, the perturbative expansion can exist in two different modes,
one in which it is directly Borel summable, while for
$g <0$ non-trivial non-perturbative contributions
are needed to get a consistent physical picture.\\

Coming back to the $D \ge 2$ field theory case, 
we stress, however, that the above
discussed Borel summability properties in our construction is
possible due to the negative (more generally complex) tail of the
perturbative expansion parameter $1/F$  in the infrared, rather than  due 
to an artificial change of sign for any $F$ values. 
One may wonder if such a sign alternation of the
badly behaving infrared factorials, may not alter the other way
round the signs of the UV renormalons, which originally have the good
(alternated) signs in AFT\cite{renormalons}. It is
easily realized that they are in fact unaffected by the infrared properties of
$F <0$, since by definition the UV renormalons originate only from the domain
$\mu \gg \Lam$ (more precisely the Borel integral equivalent to Eq.~(\ref{BI})
for UV renormalons corresponds to integration from $\mu^2 < q^2 <\infty$). In
this range, $F$ is necessarily real positive and large, see
Fig.~\ref{lambert} and Fig.~\ref{Fplot}. \\
Remark finally that, only from the properties of $F$ around $F \lsim 0$,  the
Borel sum in (\ref{directBS})  reproduces qualitatively
the asymptotic behaviour of the {\em exact} 
$1/N$ result Eq.~(\ref{MexactEi}) in the $O(N)$ GN model (with $4\pi b_0
=2N-2$), except for finite terms $\gamma_E$ etc, which 
not surprisingly cannot be guessed
by our simple Borel summation of the (leading) renormalon asymptotic behaviour
in Eq.~(\ref{BI}), and with only the first order RG dependence included.
Thus, at this stage the Borel summability property of the series for $F <0$
plays a rather formal role, since the leading order
Borel sum Eq.~(\ref{directBS}) is not expected to be a good numerical
approximation of the exact mass gap. (We
shall see in section \ref{Num} that there are more efficient
approximations to the exact mass gap.)\\

It is not difficult to work out the exact second RG order generalization
of Eq.~(\ref{directBS}). 
For this purpose, it is convenient to define another change of scheme,
by a ``non-perturbative" redefinition
of $F$:
\be
F+C \equiv f
\label{NPRS2}
\ee
perturbatively equivalent to 
\be
g^2 \to \tilde g^2 = g^2\;(1 +\frac{b_1}{b_0}g^2)^{-1}\;.
\label{npscheme}
\ee
This redefinition is again motivated from the fact that
in the GN model, the RG coefficient $C <0$ in
Eq.(\ref{ABCdef}), so
that the extrapolation from the perturbative range of (\ref{MRG}) down to
$\hat m \to 0$ reaches first $F=-C$, as explained before(see
Fig.~\ref{Fplot}).  The asymptotic behaviour
of the perturbative coefficient (exact in the scheme (\ref{betuni}))
is
\be
d_{n+1}
\raisebox{-0.4cm}{~\shortstack{ $\sim$ \\ $n\to\infty$}}
(2b_0)^n\: \Gamma[n+1+C]\;
\label{irren2} 
\ee  
for $n\to\infty$. 
The corresponding Borel
integral
reads
\be
BI(M^P)[f] =  \Lam\;(2e)^{-C} e^f f^{-C} \; 
[1+\frac{\Gamma[1+C]}{4\pi b_0\,f}\int^\infty_0 \! dt e^{-t}\;
(1 -t/f)^{-1-C} \:] 
\label{BI2}
\ee
where we used the appropriate RS change 
Eq.~(\ref{NPRS2}) in Eq~(\ref{MRGn}).  So for $F < -C$,
i.e. $f <0$, we obtain 
\be
\ds
BI(M^P)[-|f|] \sim \Lam\;(2e)^{-C} e^{-|f|} (-|f|)^{-C}\;   
[1-\frac{1}{4\pi\, b_0}\:\Gamma[1+C]\:
e^{|f|}\:|f|^C\:\Gamma[-C,|f|]\:] 
\label{directBS2}
\ee
where the $1$ in the bracket refers to the pure RG part,
while the remaining part in 
Eq.~(\ref{directBS2})
resums the purely perturbative contributions. 
The latter resummation of the perturbative expansion part is 
convergent, giving a finite contribution  $\forall |f| \neq
\infty$. In particular it gives a finite result even for $|f|\to 0$:
\be
\mbox{BI[pert. series]}
\raisebox{-0.4cm}{~\shortstack{ $\to$ \\ $|f|\to 0$}}
-(-1)^{-C}\:(2e)^{-C}\:\frac{1}{4\pi\, b_0}\:\Gamma[1+C]\: 
\Gamma[-C]
\label{RG2Bsum}
\ee
which accordingly, corresponds to reach the massless limit $\hat m \to 0$
along any of the branch of $F$ below $F=-C$, see Fig.~\ref{Fplot}. Remark
also finally that Eqs.~(\ref{directBS2}), (\ref{RG2Bsum}) only depend on the
universal RG coefficient $C$ in (\ref{ABCdef}) (noting also that
$4\pi b_0 =-C^{-1}=2N-2$ in the GN model).
\section{Variationally improved mass expansion}  
\begin{figure}[htb]
\begin{center}
\mbox{
\psfig{figure=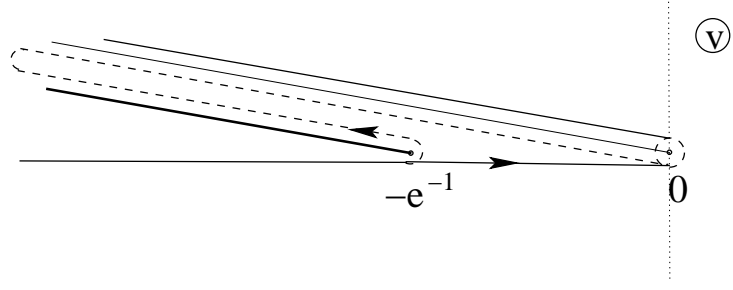,width=10cm}}
\end{center}
\caption[]{\label{contour} 
Singularities and equivalent integration contours in the
$v$ plane, for $A=1$.}  
\end{figure}

\setcounter{equation}{0}
We examine now how to complement the
above construction, based only on RG fixed point properties,
by combining it
with a specific variant of the delta-expansion, or variationally
improved perturbation (DE-VIP) method.
The latter is usually\cite{pms,delta} 
applied more directly on the
ordinary perturbative series in the coupling $g$. But the present
construction and the DE-VIP idea
are closely related, the DE-VIP being essentially a
reorganization  of the interaction terms of the Lagrangian,  
with the introduction of a
trial mass parameter, for physical quantities relevant in
the massless limit of the theory. We will see that
the DE-VIP can give further improved Borel convergence properties. 
We define the (linear) DE  as the power series obtained formally after the
substitution 
\be 
m(\mu) \to (1-\delta)\;m_v; \; g^2(\mu) \to  \delta\:g^2(\mu) 
\label{substitution}
\ee  
within any perturbative expressions at arbitrary order, where 
$m(\mu)$ is the renormalized Lagrangian mass 
(in e.g. $\overline{\small{MS}}$ scheme),
$\delta$ the new expansion parameter,
and $m_v$ an {\em arbitrary} adjustable mass. 
(\ref{substitution}) is equivalent to 
adding and subtracting to the massless Lagrangian a ``trial" 
mass term $m_v$ [$\delta$ interpolating between the 
free ($\delta=0$) and the
interacting  {\em massless} Lagrangian ($\delta=1$)], 
and is entirely
compatible with renormalization\cite{gn2} and  
gauge-invariance\cite{qcd2}.     
After substitution (\ref{substitution}),
\be  
M^{P}(\hat m, \delta) \equiv \sum_k a_k(\hat m) \delta^k
\ee
can be most
conveniently directly resummed,
for $\delta\to 1$,
by contour integration\cite{gn2} around $\delta=0$, to arbitrary order $K$:
an appropriate change of variable
allowing to study the $m(\mu)\to 0$ (equivalently  
$\delta \to 1$) limit in Eq.~(\ref{substitution}) is: 
\be
\delta \equiv  1-v/K \;;\;\;\;    m_v =  K^\gamma \;\hat m_v\;.
\label{rescale}
\ee
Eq.~(\ref{rescale}) is simply a convenient
way of parameterizing how rapidly the Lagrangian mass $ m(\mu) \to 0$ limit
is reached (as controlled by $\gamma\le 1$) as function
of the (maximal) delta-expansion order $K$. Similarly to
refs.~\cite{deltaconv,deltac} the point is to adjust 
the rates at which 
$m(\mu) \to 0$ ($\delta\to 1$) and $K\to\infty$ are 
simultaneously reached, with no a priori need of invoking explicit
PMS optimization principle.
The final contour integral summation takes a simple
form, for $K\to\infty$:
\be
\ds
\frac{M^P}{\Lam} \sim 2^{-C}\frac{1}{2\pi i}\oint dv \;
e^{(v/m^{\prime\prime})}\;F^{-A}[v]\:(C+F[v])^{-B}[1+\frac{1}{2
b_0\,F[v]}\sum^N_{n=0} \frac{d_n}{F^n[v]}]
\label{contour1}
\ee 
where $m^{\prime\prime} \equiv \hat m_v/\Lam$, $N$ is the maximal perturbative order, and
after a deformation the contour encircles the semi-axis $Re[v]<0$ 
(see Fig. \ref{contour}).
%%%%% FIGURE  3 %%%%
For simplicity we fix from now the
scaling parameter in Eq.~(\ref{rescale}) 
to its maximal value ($\gamma =1$) still compatible
with massless limit (i.e. $m_v \to 0$). (NB The
general $\gamma$ scaling (\ref{rescale})
can be analyzed in a way more similar to the 
oscillator~\cite{deltaconv,deltac}, i.e. without the peculiar
contour $\delta$-summation Eq.~(\ref{contour1}), see Appendix D.   
But this gives rather cumbersome algebraic and numerical
analysis.)
Eq.~(\ref{contour1}) can be well
approximated analytically (at least for slightly restricted RS
choices, as will be indicated):
\be
M^P/\Lam \sim
1 +\frac{1}{4\pi b_0}\;
\sum^N_{q=1} \left[\; \sum^{N-q}_{p=0} \frac{\Gamma[p+q] 
(p+q+A)(q+A)^{p-1}}{A^p\:\Gamma[1+p]\: \Gamma[1+q/A]}\; \right]
(m^{\prime\prime})^{-q/A}  \label{Mpolesum}
\ee
where we used Eq.~(\ref{Fexp}), leading to the (exact) expression
\be
F^{-n-A}(m_v\;v ) = \sum^\infty_{p=0} \frac{(n+A)(n+A-p)^{p-1}}{p!\;A^p}
(m_v\;v)^{-1 +\frac{p-n}{A}}
\label{F-nexp}
\ee
together with
\be
\frac{1}{2\pi\,i}\oint dv e^v v^z = \frac{1}{\Gamma[-z]}\;,
\ee
and also
assumed the leading
renormalon behaviour\footnote{The original $n!$ 
coefficients in Eq.~(\ref{irren}) correspond
to $\Gamma[p+q]$ in (\ref{Mpolesum}). Higher order refinements on infrared
renormalon structure may easily be implemented: it essentially
replaces $(n-1)! \to  \Gamma[n +C](1+r_1(scheme)/n+..)$ where 
 $r_1$ depends on RS~\cite{renormalons}, without affecting the
convergence properties discussed below, cf. Eq.~(\ref{irren2}).}
Eq.~(\ref{irren}). In Eq.~(\ref{Mpolesum}) we also made
a convenient reshuffling of summation indices, $n \to p+q$, where
$n$ is the original perturbation order, $p$ is the order of the expansion
in Eq.~(\ref{Fexp}), and $q$ is the order of the (resulting) expansion
in $(m^{\prime\prime})^{-q/A}$. \\
In fact, some restrictions apply\cite{KRlet} to  
Eq.~(\ref{Mpolesum}): the sum over $p$ is bounded as given, iff
\be
1/A \in \mathbf{N}^*\;,
\label{Ainvint}
\ee
since $1/\Gamma[1+(n-p)/A] =0$ for  $p \ge n+1$,
which we assume in this section for simplicity. (This is not
much restrictive, except that for arbitrary AFT it is generally 
not possible both that
$A$ satisfies (\ref{Ainvint}) {\em and}
$B=0$ in Eq.~(\ref{ABCdef}), as assumed in
(\ref{contour1}). But the more general scheme
$B\neq 0$ simply makes Eq.~(\ref{Mpolesum}) 
algebraically more involved, without affecting
the large order behaviour and convergence properties.)\\ 
Second, strictly (\ref{Mpolesum}) is
valid only asymptotically, for sufficiently
large $N$: due to the finite convergence radius 
of expansion (\ref{Fexp}), interchanging
the sum in (\ref{Fexp}) and integration in (\ref{contour1}) is not
rigorously justified. 
However, when (\ref{Ainvint}) holds, the formerly branch point $v=0$ is 
simply a pole, which allows to choose an equivalent 
contour of arbitrarily small radius around $v=0$, thus  
always inside the convergence radius of
(\ref{Fexp}) (see the dashed small circle
contour in Fig.~3). Using Eq.~(\ref{F-nexp})
(exact for $|m_v \;v| < e^{-A} A^A$);
and $e^v = \sum_k v^k/k!$, it is easily seen that the simple poles
are for
\be 
k -(1+(n-p)/A) = -1
\ee
for which the contribution to the contour integral will be the 
coefficient  (\ref{F-nexp}), divided by $k!$: this give 
$\Gamma[1+(n-p)/A] =\Gamma[1+q/A] $. So, 
only the simple pole terms $v^{-1}$ contribute to 
Eq.~(\ref{contour1}), which sum up to give 
Eq.~(\ref{Mpolesum}) again.  
The extra contribution (around the cut at $v =
-e^{-1}$, e.g. for $A=1$)
gives the difference between the ``exact"
integral (\ref{contour1}) and expansion (\ref{Mpolesum}), 
and can be evaluated numerically. These 
contributions are easily
shown for $A=1$  to contribute as
${\cal O}( e^{-(e\:m^{\prime\prime})^{-1}}) h[N]$ 
relative to (\ref{Mpolesum}),
where $h[N]$ rapidly 
decreases for $N\to\infty$. In Table \ref{numcont} we compare, for the first 20
values of the perturbative order $n$, and for $m^{\prime\prime}=1$, the exact
integrals: 
\be
\ds
I^{exact}_n = \frac{1}{2\pi\,i}\oint dv e^{v/m^{\prime\prime}} F^{-n-A}[v]
\label{Iexact}
\ee
which we calculated numerically with Mathematica\cite{matha}, 
with the approximation using the expansion (\ref{F-nexp}) 
(thus neglecting the extra contribution around $-e^{-1}$):
\be
\ds
I^{series}_n = \sum^n_{p=0} \frac{(n+A)(n+A-p)^{p-1}}{p!\;A^p
\:\Gamma[1+(n-p)/A]}
(m_v)^{-1 +(p-n)/A}
\label{Iseries}
\ee
that leads to Eq.~(\ref{Mpolesum}). We also give 
in the third column of Table \ref{numcont} the extra $e^{-1}$ cut contribution,
evaluated independently numerically for consistency.\\ 
\begin{table}
\begin{center}
\begin{tabular}{|l| |c|c|c|}
\hline
n  & $I^{exact}_n(m^{\prime\prime}=1)$  & $I^{series}_n(m^{\prime\prime}=1)$
& -$e^{-1}$ cut contribution\\ \hline
0  & 1.137545624095354 & 1. & 0.1375456240953545\\
\hline
2  & 4.992495983527079 & 5. & -0.00750401647292076\\
\hline
4  & 8.165946096293188 &  8.166666666666666 & -0.0007205703734776136 \\
\hline
6  &  8.780755015889008 & 8.780555555555555 & 0.0001994603334534161\\
\hline
8  & 7.28191740833204 & 7.281944444444444 & -0.00002703611240395531\\
\hline
10  &  5.01821660133935 & 5.0182137345679  & 2.86677144956 10$^{-6}$ \\
\hline
12  &  2.99921746894628 &  2.999217731214259 & -2.62267978889 10$^{-7}$ \\
\hline
14  &  1.597938546376994 &  1.597938525158887 & 2.121810749323 10$^{-8}$ \\
\hline
16  &  0.773538637086908 & 0.773538638583302 & -1.4963940087 10$^{-9}$ \\
\hline
18 & 0.3450035964997671 & 0.3450035964141139 & 8.5653262261 10$^{-11}$ \\
\hline
20 &  0.1432800076663147 & 0.1432800076691064 & -2.79173906215 10$^{-12}$ \\
\hline
\end{tabular}
\vspace{5mm}
\end{center}
\caption[]{\label{numcont} Comparison between Eq.~(\ref{Iexact})
(evaluated numerically), Eq.~(\ref{Iseries}), and the extra cut
contributions.}
\end{table}
%%%%%%%%%%%%%%%
When $m^{\prime\prime}\to 0$, the discrepancy between the exact integral
and the analytical resummation in Table \ref{numcont} decreases rapidly, as
expected, even for small $n \to 0$ (but the numerical integration becomes
unstable for very small $m^{\prime\prime}$). 
If $A\neq 1$ and 
$B$ arbitrary, contributions
from extra cuts are not so
simply estimated, and we were only able 
to check numerically
that they are negligible with respect to (\ref{Mpolesum})
for sufficiently large $N$.] Thus for 
large enough $N$ (and/or small $m^{\prime\prime}$) 
those contributions are unessential for the convergence
properties discussed below.\\
     
The factorial damping of 
coefficients,
as compared
to the original perturbative expansion, is explicit in 
Eq.~(\ref{Mpolesum}). Yet,  the damping is insufficient to make this series
for $N \to \infty$ readily convergent. 
For any low $p \ll N$, renormalon
factorials are overcompensated if $A \le 1$,
but the $\Gamma[1+q/A]$ damping decreases in strength
as $p$ increases, giving increasing contributions to the sum over $p$. 
The leading contributions 
to the coefficients of (\ref{Mpolesum}) happen at intermediate values of
$p$.
Nevertheless, the idea of damping factorials from appropriate
RS choice does survive, when considering the Borel transform
of Eq.~(\ref{Mpolesum}), as examined next. 
\section{Borel convergence of DE-VIP}
\setcounter{equation}{0}
We are now ready to combine the previous DE-VIP behaviour of the
series in $\hat m/\Lam$ with the general
Borel convergence properties 
for $Re[F] <0$
of the initial perturbative series in $1/F$, which were examined in section 4.
For completeness we consider both a linear and non-linear version of the DE-VIP
construction, also for reasons that will be clear below. We will see that the
DE-VIP expansion can generally improve (accelerate) 
the Borel convergence properties of the $1/F$ expansion.
\subsection{Linear method}
For any given choice of contour avoiding the pole in the Borel plane $t$
(or cut at higher RG order, see Eq.~(\ref{BI2})), 
we can apply the DE-VIP as defined in section 4,
introducing the $\delta$--expansion and contour resummation as
in Eq.~(\ref{contour1}), 
but now on the Borel integral Eq.~(\ref{BI}) (or Eq.~(\ref{BI2})).
It leads to: 
\be
\tilde M^P_{var}(\hat m_v) \sim 
\frac{2^{-C }}{2i\pi} \oint dv e^{v}\;
\frac{\hat m_v}{F^A\,(C+F)^B}\;\int^\infty_0 dt e^{-t}\;
[1 +\frac{1}{4\pi\,b_0\,F}(1-t/F)^{-1} \:]
\label{BIvar}
\ee
where $F\equiv F[\hat m\:v]$ and the integrand is to be be understood as its
formal expansion in  $t/F[\hat m\: v]$.
Interchanging the contour and Borel integrals, 
the Borel transform integrand, which is a function of $t$, is essentially
Eq.~(\ref{Mpolesum}) but with the replacement
\be
\Gamma[p+q] \to t^{p+q}
\ee
standard from the Borel transform, except that in our case 
we did some reshuffling of summation indices, as already indicated
after Eq.~(\ref{Mpolesum}).
One can find
after some algebra the asymptotic behaviour for (the original
maximal perturbative order) 
$N\to \infty$:
\be
\displaystyle
\tilde M^P_{var}(m^{\prime\prime}) \sim const.\;\Lam \;[ 1+  
\int^\infty_0 \!\frac{dt}{4\pi\,b_0}\,\,
\sum^\infty_q 
\frac{(t^A e^t/m^{\prime\prime})^{q/A}}{\Gamma[1+q/A]}\;]
\label{BIvarbis}
\ee
where  we work again here for simplicity at first RG order, neglecting 
the pure RG resummed $(C+F)^{-B}$ term in (\ref{BI}). 
It thus appears that the asymptotic behaviour of the Borel integrand 
in Eq.~(\ref{BIvarbis}) is that of an entire series (at least for $A>0$), i.e.
with no poles for $0< t < \infty$. More precisely, the pole at $t_0=1$ in the 
original (standard) Borel integrand has been pushed to $t_0 
\to +\infty$ due to the factorial damping, so that the Borel integral is 
no longer ambiguous. However, integral 
(\ref{BIvarbis}) is badly divergent at $t\to\infty$, at least for
$Re[m^{\prime\prime}]>0$, so that the  series is not Borel summable for
standard (perturbative) $m^{\prime\prime}$ values. In fact it is important to
remark that this non convergent result is obtained when considering the exact
expansion of $F$, Eq.~(\ref{Fexp}), to all orders. Naively, one may have
thought that the contour integral would be dominated for $q \sim N \gg p$, by
the apparently "leading" terms for $m^{\prime\prime} v \simeq 0$ in 
Eq.(\ref{contour1})\footnote{Also by simple
analogy with the oscillator, which has a series in $g/m^3$, cf.
Eq.~(\ref{E0exp1}), thus formally identical to taking $p=0$ in
Eq.(\ref{Fexp}),(\ref{Mpolesum}), and the replacements $A \to 2/3$,
$m^{\prime\prime}\to m^2 g^{-2/3}$.}. But the asymptotic behaviour of our
complete series Eq.(\ref{Mpolesum}), as well as its Borel transform 
Eq.~(\ref{BIvarbis}), appear much less
intuitive than e.g. the oscillator energy levels expansions.
Therefore, it is due to the ``late" terms of expansion
(\ref{Fexp}) (corresponding to the terms $p \sim N-q$ when $N\to\infty$
in Eq.~(\ref{Mpolesum})), that the Borel integral ultimately diverges.
We have checked by direct numerical contour integration of the Lambert
function, which can be done e.g. with Mathematica\cite{matha}, that as
$N\to\infty$ expression (\ref{Mpolesum}) is asymptotically correct (see
Table \ref{numcont}), while a finite truncation of (\ref{Fexp})
to the first few terms would not be a good approximation.\\

Now conversely, the integral in Eq.~(\ref{BIvarbis}) can converge,
for $Re[m^{\prime\prime}]<0$. This is the case at least for $A=1$, which can
always be chosen by an appropriate and simple RS change, according to
Eqs~(\ref{RSC}). (In particular, such RS change only affect the very first
ordinary perturbative order coefficients (see Appendix C), thus cannot modify
their behaviour at large orders, Eq.~(\ref{irren})).   
Now, since $m^{\prime\prime}\equiv m_v/\Lam$ is an arbitrary parameter, 
it should be legitimate to reach the chiral limit $m^{\prime\prime}\to 0$, of
main interest here, within the Borel-convergent half-plane
$Re[m^{\prime\prime}]<0$.  For $A\ne 1$, one may also
choose the arbitrary parameter $m^{\prime\prime}$ with
$Re[(m^{\prime\prime})^{1/A}] <0$ such that (\ref{BIvarbis})
converges, though this appears not always possible for any arbitrary $A$
values.  Yet, the general Borel convergence properties obtained in section 4
prior to the DE-VIP transformation of perturbative expansion, was valid
independently of $A$ values. In fact, the scheme choice limitation
in obtaining convergence of Eq.~(\ref{BIvarbis}) is 
only an artifact of our simplest choice of the $\delta$-expansion
summation defining the DE-VIP series and leading to (\ref{BIvarbis}), as
examined in next sub-section.  
\subsection{Non-linear variational expansion}
We consider now a convenient modification of our variational expansion method,
defining a non-linear transformation in the 
variational parameter $m_v$,
as compared to 
Eq.~(\ref{substitution}). First we
remind that Eq.~(\ref{Fdef}): \be
F = \ln \frac{\hat m_v}{\Lam} -A \ln F
\label{Fdef2}
\ee
defines $F$ as a systematic power
expansion in $(m_v/\Lam)^{1/A}$, cf. Eq.~(\ref{Fexp}), rather than a single
power of $m_v/\Lam$, as reminiscent of the renormalization logarithms. This is
the main difference (and main source of complexity) with the simpler oscillator
energy levels, for which each perturbative expansion order is a simple
$(g/m^3)^n$ power\cite{ao}.  Now, since we introduce a 
reparameterization of the Lagrangian interaction terms, 
cf. Eq.~(\ref{substitution}),
it may be possible to remove
the logarithm from $F$ by an appropriate definition of the alternative
interaction terms. Heuristically, this can be achieved by redefinition of the
arbitrary mass parameter, for example: 
\be
\ds
\hat m_v \to \hat m_v\; e^{\hat m_v^{1/A}} \equiv \hat m^\prime_v
\ee
which immediately gives, replacing in Eq.~(\ref{Fdef2}): 
\be
F = (\hat m^\prime_v)^{1/A}
\ee
i.e. a {\em single} power dependence on $\hat m^\prime_v$. Of course, the
above manipulation is just an equivalent change of variable: the theory is
completely equivalent in terms of $m^\prime_v$, since all the complexity
of (\ref{Fdef2}) is now hidden in the relation between $m_v$ and $m^\prime_v$.
Nevertheless, if one can make such a transformation to depend
on the DE-VIP expansion parameter $\delta$ in Eq.~(\ref{substitution}),
the contour integrand in Eq.~(\ref{contour1}) resumming the DE-VIP
expansion can have a simpler
$v$-dependence.  More precisely, instead of the linear DE defined by 
Eq.~(\ref{substitution}), we consider in the  (variationally modified)
Lagrangian a mass term
\be
(1-\delta)\: m_v \:{\cal G}(m_v,\delta) \;\bar \Psi \Psi
\ee
where ${\cal G}$ now explicitly depend on the new DE-VIP expansion
parameter $\delta$, thus the non-linearity. ${\cal G}(m_v,\delta)$
is only required to be a scale (RG)
invariant function (so that it does not affect any of the RG properties) and to
have a $\delta$ dependence only constrained by requiring that it
still interpolates between the free massive Lagrangian (for $\delta\to 0$) and
the interacting massless theory (for $\delta \to 1$),
the original theory. The function ${\cal
G}$ thus defines a (non-linear) interpolation function which is
otherwise essentially arbitrary.  This may be viewed as a
particular "order dependent mapping" (ODM)\cite{odm}.\\
Specifically we will use here a convenient\footnote{The
choice of the non-linear interpolation is clearly not unique. Other
choices\cite{damthese} differ in the explicit form of the resulting DE-VIP
expansion, but have similar asymptotic properties at large perturbative
orders.}  form for ${\cal G}$: 
\be
\hat m_v/\Lam \equiv m^{\prime\prime} \to m^\prime {\cal G}(m^{\prime\prime},v)
\label{mtomprime}
\ee
\be
{\cal G} = e^F F^{A-1} (C+F)^{B-C}
\ee
with $F\equiv F[m^{\prime\prime}\:v]$,
i.e. a deformation of the mass term depending also
on the RG parameters. This gives
\be
F \equiv \ln(m^{\prime\prime} v) - A \ln F -(B-C)\ln(C+F)
= \ln (m^\prime v) +F - \ln F 
\ee
\be
\Leftrightarrow \hspace{2cm} F \equiv (m^\prime v)  
\label{Fnlform}
\ee
With the simple power form of $F$ in Eq.~(\ref{Fnlform}), 
the contour integral Eq.~(\ref{contour1}) now reads:
\be
\ds
\frac{M^P}{\Lam} \sim 2^{-C}\frac{1}{2\pi\,i}
\oint \frac{dv}{v} e^{v\,(1+m^\prime)} (C+m^\prime\,v)^{-C} 
[1+\frac{1}{4\pi\,b_0\,m^\prime\:v} \sum^N_{n=0} \frac{d_n}{(m^\prime\:v)^n}\;]
\ee
which is immediately integrable (again neglecting here the higher RG
order $C$ dependence for simplicity):
\be
\frac{M^P}{\Lam} \sim 1+\frac{1}{4\pi\,b_0}\sum^\infty_{n=0}
\frac{d_{n+1}}{\Gamma[n+2]}\;(\frac{m^\prime+1}{m^\prime})^{n+1} 
\label{nlsum}
\ee
which is now valid for arbitrary values of the RS parameter $A$.
With $d_n \sim n!$ for $n\to\infty$ 
the series (\ref{nlsum}) thus converges iff
\be
|\frac{1+m^\prime}{m^\prime}| < 1
\label{discnl}
\ee
which is only possible if $-\infty < Re[m^\prime] <-1/2$, 
thus for negative values of the arbitrary mass $m^\prime$, in consistency
with the linear method results in previous sections. Taking more
specifically the asymptotic behaviour of the perturbative coefficients
Eq.~(\ref{irren}) at first RG order, the series can be resummed to
\be
\frac{M^P}{\Lam} \sim 1-\frac{1}{4\pi\,b_0}
\;\ln[1-\frac{m^\prime+1}{m^\prime}]\;.
\label{nlresum}
\ee
In summary, we obtain  from this non-linear method a directly convergent
series, where the convergence domain is in the $Re[m^\prime]<0 $ range of the  
trial mass parameter, and this result is now independent of the $A$ values,
as expected.
The series in Eq.~(\ref{nlsum}) may also be 
straightforwardly Borel resummed: this gives
\be
\ds
M^P \sim \Lam \;[1-\frac{1}{4\pi\,b_0}
\;\int^\infty_0 \frac{dt}{t} (e^{-t(1-r)} -e^{-t})\;]
\label{BInl}
\ee
with $r\equiv (m^\prime+1)/m^\prime$,
which converges to Eq.~(\ref{nlresum}) for $Re[r]<1$, thus for
$Re[m^\prime] <0$. Since the series in Eq.~(\ref{nlsum}) was already convergent
in the $m^\prime$ domain given by Eq.~(\ref{discnl}), the
role of the Borel summation in this case is simply that it can extend the
convergence domain, namely from (\ref{discnl}) to the full 
half-plane $Re[m^\prime] <0$.  Another advantage of such non-linear 
transformations is that the extra singularities, at 
$ v = e^{-A}(-A)^A$, are absent in this picture (in fact, they have not
completely
disappeared, no more than the RS $A$--dependence, but all this is 
hidden in the relationship between $m$ and $m^\prime$
Eq.~(\ref{mtomprime}), by definition). 
The numerical application of this non-linear method will be
illustrated in the end of the next section. 
\section{Numerical results}\label{Num}
\setcounter{equation}{0}
We are now in a position to analyze and compare in some details the numerical
results obtained from different possible approximations to the $O(2N)$ GN model
mass gap, defined from our construction.  
The Borel (or direct) convergence properties as obtained 
in sections 5 and 6 are encouraging but unfortunately only formal
properties of the large orders, asymptotic behaviour of the true series. 
For instance, the leading order Borel sum
Eq.~(\ref{directBS}) is not expected to be a good numerical approximation
of the exact mass gap, in analogy with the Borel summable
oscillator case, where the direct Borel sum expression
(see e.g. Eq.~(\ref{aoBsum}) in Appendix B),
is of not much practical use for an accurate numerical determination
of the energy levels. (Apart from a direct numerical solution of the
Schr\"odinger equation, the most efficient analytical methods to determine the
oscillator energy levels accurately are either Pad\'e approximant
techniques\cite{aoBsum}, or the ODM \cite{odm} or related optimized
delta-expansion \cite{deltaconv,deltac} methods). In $D \ge 2$ field theories
it is even less expected that the first few perturbative coefficients
(recalling that only the first two are exactly known in the GN model) are
close to the asymptotic behaviour. The precise numerical values of the first
perturbative orders could thus be a priori of much relevance for a precise
determination of the mass gap, in particular for low $N$ values. 
\subsection{Direct mass optimization} 
We start in this first sub-section by exploring a  numerical approximation
more directly motivated by the usual DE-VIP or related ``principle of minimal
sensitivity" (PMS)\cite{pms} ideas. Since the main parameter in our
construction is the arbitrary mass $\hat m_v$, and the exact result is in the
massless Lagrangian limit,  thus independent of the mass,
the PMS leads naturally to
optimize our expressions for the mass gap, e.g. (\ref{contour1}), with
respect to this trial mass parameter. We can compare these optimization
results when taking successive orders of the original perturbative expansion
into account. 
Numerical optimization results are
summarized in Tables \ref{tabopt1} and \ref{tabopt2}. More precisely,
in Table \ref{tabopt1} we look for extrema in $\hat m/\Lam \equiv
m^{\prime\prime}$ of the original mass gap expression Eq.~(\ref{MRGn}),
truncated to first and second perturbative orders, respectively, in the $\MS$
scheme. In other words, only the exactly know perturbative information is
taken into account in these results. (Note in particular that none of the
above discussed large order, resummation, and eventual Borel
convergence properties are taken into account in any way here.)
In practice we rather use the change of renormalization scheme
as explained in section 2, which leads to Eqs.~(\ref{MRGNsc})--(\ref{dnGN}),
more appropriate to the GN model. As one can see, there
exist optima in $m^{\prime\prime}$ for all cases studied, but the optimal
$M^P/\Lam$ mass gap results are rather far away from the exact ones, except
when $N\to\infty$ (where the correct result $M^P/\Lam =1$ is always recovered).
Moreover, when adding more perturbative coefficients, there is even a
substantial degradation in these optimal results. This may appear a
bit surprising, if comparing with the excellent results obtained from a
similar optimization with respect to the Lagrangian mass of the
oscillator energy levels\cite{delta}. In our opinion it simply
reflects that in the more complicated field theory case, a naive application
of PMS ideas may not always work so well, in view of the  numerous problems
that afflict the original perturbative series, as emphasized in previous
sections. \\ %%%%%%%%%%%%%%%%
\begin{table}[htb] \centering
\begin{tabular}{||l||l|l|l||}  \hline
$N$ ($O(2N)$ model) & exact & order 1 (pure RG) & order 2     
   \\ \hline 
2   & 1.8604  & 2.215  &  3.28 \\ \hline 
3  & 1.4819  & 1.730 & 2.63 \\ \hline 
5 & 1.2367 &1.464 &  2.11 \\ \hline
8 & 1.133 & 1.337 &  1.84  \\ \hline
$\infty$  & 1 & $\sim$ 1 & $\sim$ 1 \\
\hline
\end{tabular}
\caption{\label{tabopt1}$M^P/\Lam$ from direct optimization of 
Eq.~(\ref{MRGn}) (truncated to first few orders, with exact perturbative
coefficients)
with respect to $m^{\prime\prime}$, in the scheme Eq.~(\ref{MRGNsc}).}
\end{table}
%%%%%%%%%%%%%%%%%%%%%%%%%   
Next, we show in Table \ref{tabopt2} the results from a similar
optimization with respect to $m^{\prime\prime}$, but this time on the DE-VIP
resummed, contour integral expression of the mass gap: this is essentially
Eq.~(\ref{contour1}), except that again only the exactly known perturbative
coefficients are taken into account, thus  truncating Eq.~(\ref{contour1}) at
first and/or second order respectively (as indicated in Table \ref{tabopt2}).  
The results are much better than those in Table \ref{tabopt1},
which may be attributed to the expected improved properties of the contour
integration resummation. In particular, the results at first order
are very close to the exact mass gap, at least for $N\ge3$. 
We observed that for low $N$ values the optima is for rather
small $m^{\prime\prime}_{opt}$ (e.g. $m^{\prime\prime}_{opt}\sim 0.087$ for
$N=2$ in  the second column of Table \ref{tabopt2}). When $N$ increases,
$m^{\prime\prime}_{opt}$ also increases, up to a maximum $m^{\prime\prime}_{opt}\sim 0.26$ 
for $N\sim 6$, and then again $m^{\prime\prime}_{opt}\to 0$ as $N\to\infty$.
In connection with our Borel convergence results of section 4
it is interesting to remark that there also exist optima for $m^{\prime\prime} <0$
values, see Table
\ref{tabmneg}.
However, inclusion of
higher perturbative orders makes 
the results to degrade rather rapidly, contrary to what could have been
expected, as one can see in Table \ref{tabopt2}.  
We conclude that such a rather naive optimization, keeping only the lowest
perturbative orders into account in direct inspiration of the PMS/DE-VIP
ideas,
is not much conclusive within our framework. Note also the excellent
optimization results obtained, if considering only the first order, i.e. the
pure RG dependence only. (In 
fact, in the scheme Eq.~(\ref{MRGNsc}), the ``first
order" in Tables  \ref{tabopt1},  \ref{tabopt2} 
only involve the pure RG dependence, since in the $O(N)$ GN model\cite{gn2}
the first perturbative order coefficient in Eq.~(\ref{MRGn}): 
$d_1(\MS)\equiv 0$, so that $\tilde d_1 = -B C$ from Eq.~(\ref{dnGN}).)
%%%%%%%%%%%%%%%%
\begin{table}[htb] \centering
\begin{tabular}{||l||l|l|l||}  \hline
$N$ ($O(2N)$ model) & exact & order 1 (pure RG) & order 2     
   \\ \hline 
2   & 1.8604  & 2.0493  &  2.85 \\ \hline 
3  & 1.4819  & 1.5156 & 2.10 \\ \hline 
5 & 1.2367 &1.257 &  1.62 \\ \hline
8 & 1.133 & 1.152 &  1.40  \\ \hline
$\infty$  & 1 & $\sim$ 1 & $\sim$ 1 \\
\hline
\end{tabular}
\caption{\label{tabopt2}$M^P/\Lam$ from direct optimization of 
Eq.~(\ref{contour1}) (truncated to first few orders, with exact perturbative
coefficients)
with respect to $m^{\prime\prime}$, in the scheme Eq.~(\ref{MRGNsc}).}
\end{table}
%%%%%%%%%%%%%%%%%%%%%%%%%  
%%%%%%%%%%%%%%%%
\begin{table}[htb] \centering
\begin{tabular}{||l||l|l|l|l|l||}  \hline
$N$ ($O(2N)$ model) & 2 & 3 & 5  & 8      & $\infty$   \\ \hline 
$M_{opt}/\Lam$   & 1.929  & 1.357 &  1.106 & 1.024  & $\sim$ 1 \\ \hline 
\end{tabular}
\caption{\label{tabmneg}$M^P/\Lam$ similarly to Table \ref{tabopt2} 
optimizing with respect to $m^{\prime\prime}$, where $m^{\prime\prime}_{opt} <0$. Only the
first order perturbative contributions are shown here.}
\end{table}
%%%%%%%%%%
\subsection{Pad\'e approximants including
$1/N^2$ contributions}\label{pade}  
The second kind of approximation that we have performed, is still not related
to the Borel convergence properties as described in the above sections, being 
again essentially based on the original basic perturbative expansion, but
treating the variational expansion in (\ref{substitution})--(\ref{contour1})
in an alternative manner to extrapolate in
the infrared, non-perturbative region. More precisely, in order
to define the massless limit $m^{\prime\prime}\to 0$ without resumming the
complete series, we have used a variation\cite{gn2} of Pad\'e
approximant\cite{pade} (PA) techniques.  The latter are know to give a
reliable resummation procedure  of perturbative expansion in various
situations, provided that the singularities of the original expansion are
controllable to some extent.
The
detailed construction of our particular PA is given in Appendix \ref{Appade}.  
These approximants have the generic form 
\be
\frac{M^P}{\Lam} = const.\; A_0 \;\exp \left[
1- \int^1_0 du \;P_{[p,q]}(u) - \int^\infty_1 du \;[P_{[p,q]}(u)-1
+\frac{C+1}{u} ] \right]
\label{padenum}
\ee
where, prior to the PA construction, the very same contour integral as the one
in Eq.~(\ref{contour1}) has been performed (see Appendix \ref{Appade}). In
Eq.~(\ref{padenum}) the overall constant takes into account any terms from e.g.
RG dependence (such as typically the $2^{-C}$ factor in Eq.~(\ref{contour1}),
or depending eventually on the
choice of renormalization scheme). The  PA functions
$P_{[p,q]}(u)$ are, as usual\cite{pade}, rational fractions
of polynomials in the relevant variable $1/u \sim 2b_0\: g^2_{eff}$, which is
related to $m^{\prime\prime}$ in Eq.~(\ref{contour1}) as follows:
\beq
 u^A e^u (C+u)^{B-C} = m^{\prime\prime} \;,
\label{udeftext}
\eeq 
and accordingly have by construction the 
property of disentangling the usual perturbative 
$\ln\:[\ln..[m^{\prime\prime}]]$ behaviour.
Standard perturbation theory corresponds to $u\to\infty$
and the massless limit correpsonds to $u \to 0$. The splitting of integration
ranges  in Eq.~(\ref{padenum}) is due to a necessary subtraction of
UV divergences, uniquely fixed by the perturbative expansion.
By construction (see Appendix \ref{Appade}) the PA in Eq.~(\ref{padenum})
have a finite, regular $u\to 0$ limit, which
 somehow selects a limited (but not unique) 
form of possible PA, for a given order of
the original perturbation.\\
In ref.~\cite{gn2}, using for the $O(2N)$
GN mass gap only the first
two orders of the original perturbative series coefficients in
Eq.~(\ref{MRGn}), we had obtained from appropriate $p, q$ orders of the
Pad\'e approximants, reasonably good numerical approximations of the exact mass
gap\cite{FNW}:
\be
\frac{M^P_{exact}(N)}{\Lam} = (4e)^{1/(2N-2)}/\Gamma[1- 1/(2N-2)]\;.
\label{Mexact}
\ee
The advantage of these PA is that we can systematically incorporate
higher orders (approximations) of the original perturbative
series. Accordingly, to check the stability, and eventual numerical convergence
in our approach, we incorporate in this analysis a definite
information on higher perturbative orders (recalling that their exact
expressions beyond two loops are still unknown in the GN model).
To this aim we 
exploit the {\em exactly
known} $1/N$ and $1/N^2$ dependence of the GN model RG coefficients, obtained
in the $\MS$ scheme
from an analysis near the perturbative critical point 
in a $2-D \equiv \epsilon$--expansion
in ref.~\cite{GraceyN2}.
The idea is that, since all the perturbative coefficients are
RS dependent (as discussed in section 2 and Appendix
\ref{ApN2}) one can transfer the
(exact) RG information of order $1/N^2$ into perturbative coefficients $d_n$
of order $n \ge 3$ simply by an appropriate RS change. This RS change is
essentially a change from the $\MS$ scheme, in which the exact 
$1/N$ and $1/N^2$ RG
dependence was obtained\cite{GraceyN2}, to the 
``two-loop truncated"
't Hooft scheme, defined in Eqs.~(\ref{betuni}), (\ref{gamuni}).
More precisely, beyond second order in an arbitrary scheme where
$b_i, \gamma_i \neq 0$ for $i \ge 2$ (such as the $\MS$ scheme),
the pure RG dependence can always be expanded in perturbation, thus taking
the form of specific contributions to the coefficients of $1/F^n$ in 
Eq.~(\ref{MRGn}) (see Appendix A for details).
Actually, this does not generate the exact $1/N^2$ perturbative
coefficients, but only an approximation of these, in a certain
scheme. In order to uniquely fix those resulting perturbative contributions,
coefficients of $1/F^n$, we also need to fix the perturbative coefficients
in the original (truncated) scheme, that we assume for simplicity to be
zero. This may be viewed again as a particular scheme choice.
Whether this assumption is a good approximation or
not can only be decided by the numerical analysis.
The precise
link with the $1/N$ and $1/N^2$ information, and how the RS change generates
perturbative coefficients is detailed in Appendix \ref{ApN2}. After some
straightforward algebra, we obtain in this way systematic 
corrections to the mass gap in the form of arbitrary order
perturbative coefficients in Eq.~(\ref{MRGn}). This gives e.g. for the first
few order terms:
\bea
& \tilde d_1 & = \frac{3}{4}\;\frac{(N-1/2)}{(N-1)^2} \nn \\
& \tilde d_2 & \simeq -0.3025 \;\frac{(N-1/2)(N-1.4708)(N-0.0127)}{(N-1)^4} \nn
\\  & \tilde d_3 & \simeq 0.9375 \;\frac{(N-1/2)(N+0.1645)
(N-0.6986)(N-1.1486)(N-1.3595)}{(N-1)^6} 
\label{padcoeff}
\eea 
etc. Including these results into the PA Eq.~(\ref{padenum})
we obtain the numerical results shown up to fourth perturbative order in 
Tables \ref{padtab} and \ref{padtab2} for two different RS choices 
below. Since  by construction those PA are defined in the massless limit, the
only remaining arbitrariness is the one due to the  scale (or scheme)
dependence\footnote{Defining the RS change from the 
$\MS$ to the truncated scheme is uniquely fixed except for one RS parameter,
see Appendix \ref{ApN2}.}.
This leads us to study also this remaining scale dependence by
looking for possible optima\cite{gn2} with respect to an arbitrary  scale
change:  
\be
\mu \to a\;\mu\;.
\label{padscale}
\ee
%%%%%%%%%%%%%%%%
\begin{table}[htb] \centering
\begin{tabular}{||l||l|l|l|l|l||}  \hline
$N$ ($O(2N)$ model) & exact & pure RG & order 2  &  order 3      &  order 4   
   \\ \hline 
2   & 1.8604  & 2.1213  & 1.6206$^\star$ [1.524]  &  2.044 [2.0434] &  2.1103
[2.097] \\ \hline 
3  & 1.4819  & 1.4865 & 1.3456$^\star$ [1.344]  & 1.5238  [1.4914] & 1.5372
[1.4986]\\ \hline 
5 & 1.2367 &1.2268 & 1.1917$^\star$ [1.186] & 1.27375 [1.2397] & 1.2733
[1.2394]\\ \hline
8 & 1.133 & 1.1258 & 1.1205$^\star$ [1.109] & 1.16518  [1.1355] &  1.1631 
[1.1344]\\ \hline
$\infty$  & 1 & 1 & 1.0165[0.9972] & 1.0165[0.9972]  & 1.0165[0.9972] \\
\hline
\end{tabular}
\caption{\label{padtab}$M^P/\Lam$ from a Pad\'e approximant $P[2,3]$, in the
scheme defined by Eqs~(\ref{MRGNsc})--(\ref{dnGN}),
(\ref{padcoeff}),(\ref{padscale}), optimized with respect to  the scale
parameter $a$. The order 2 values indicated as $\star$, were obtained in
ref.\cite{gn2}. Also, the (unoptimized) $\overline{MS}$ values (i.e. $a=1$)
are indicated in brackets.} \label{tabpad}
\end{table}
%%%%%%%%%%%%%%%%%%%%%%%%%
From Tables \ref{padtab} and \ref{padtab2} we can see
that there are indeed always optima in $a$, and we observed that, as the order
increases, these optima become relatively flat, and are 
closer to the original $\MS$
scheme (corresponding to $a=1$). 
Those properties are empirically quite
satisfactory, as one expects on general grounds that the approximations
should be less and less sensitive to the scale (or scheme), since the exact
result does not depend on the latter. 
Moreover, the comparison of the second, third, and fourth order 
indicates  a reasonable numerical agreement with the exact results. Indeed the
third and fourth orders in the $\MS$ scheme [the numbers in brackets] are very
close to the exact results, at least for $N \ge 3$. \\ 
%%%%%%%%%%%%%%%%
\begin{table}[htb] \centering
\begin{tabular}{||l||l|l|l|l|l||}  \hline
$N$ ($O(2N)$ model) & exact & pure RG & order 2  &  order 3      &  order 4   
   \\ \hline 
2   & 1.8604  & 2.1213  & 1.872$^\star$ [1.707]  &  2.079 [2.057] &  2.048
[2.008] \\ \hline 
3  & 1.4819  & 1.4865 & 1.487$^\star$ [1.486]  & 1.528  [1.519] & 1.510
[1.504]\\ \hline 
5 & 1.2367 &1.2268 & 1.265$^\star$ [1.252] & 1.270 [1.251] & 1.262
[1.245]\\ \hline
8 & 1.133 & 1.1258 & 1.163$^\star$ [1.145] & 1.162  [1.140] &  1.157 
[1.137]\\ \hline
$\infty$  & 1 & 1 & 1.0165[0.9972] & 1.0165[0.9972]  & 1.0165[0.9972] \\
\hline
\end{tabular}
\caption{\label{padtab2}$M^P/\Lam$ from a Pad\'e approximant $P[2,3]$, in the
original scheme corresponding to Eq.~(\ref{MRGn}). Otherwise same captions as
for table \ref{padtab}.}
\end{table}
%%%%%%%%%%%%%%%%%%%%%%%%%
We also give in Table \ref{pert1tab}, for indication, the numerical values of
the  perturbative coefficients up to fourth order that enter the PA analysis, 
for both schemes corresponding to the results in Table  \ref{padtab}
[and \ref{padtab2}], respectively. One can see for instance that the
perturbative coefficients
for $N=2$ are increasing as the perturbative order increases, while it is
the reverse for $N\ge 3$, which may explain why the results become better
for $N\ge3$.\\ 
A more curious
and perhaps rather remarkable fact, is that the PA results are also very close
to the exact mass gap, when considering {\em only} the pure RG dependence: i.e.
neglecting all perturbative orders, in which case the PA in
Eq.~(\ref{padenum}), after the appropriate subtraction of divergences (see
Appendix \ref{Appade}), essentially reduces to a constant depending only on
the RG parameters A, B, C. This is indicated as ``pure RG" in the
third column of Tables \ref{padtab} and \ref{padtab2}. If taking those PA
results at face value, it seems that the pure RG approximation gives very good
results, for $N \ge 3$, then there is some degradation when only the lowest
perturbative orders are included, and then results become again closer to the
exact ones when more perturbative orders are included. At this
level, we cannot completely exclude that this behaviour maybe a numerical
accident, though this seems unlikely, as the comparison of two different
schemes in Tables \ref{padtab} and \ref{padtab2} show a definite stability of
these results.\\
%%%%%%%%%%%%%%%%
\begin{table}[htb] \centering
\begin{tabular}{||l||l|l|l|l||}  \hline
$N$ ($O(2N)$ model) & $\frac{d_1}{2b_0}$ & $\frac{d_2}{(2b_0)^2}$  
&  $\frac{d_3}{(2b_0)^3}$  
&  $\frac{d_4}{(2b_0)^4}$       \\ \hline 
2    & 1.125 [0.375] & -0.477 [0.085]  &  2.16 [1.514] &  6.39 [-2.98] \\ 
\hline 
3    & 0.469 [0.156]& -0.216 [0.077]  & 0.810  [0.439] & 0.508 [-0.698]\\ 
\hline 
5 &0.211 [0.07] & -0.094 [0.045] & 0.321 [0.158] & -0.012 [-0.211]\\ \hline
8 & 0.115 [0.038] & -0.049 [0.027] & 0.162  [0.078] & -0.04 [-0.093]\\ \hline
$\infty$  & 0 & 0 & 0  & 0 \\
\hline
\end{tabular}
\caption{\label{pert1tab}$M^P/\Lam$ perturbative coefficients of $1/F$,
Eq.~(\ref{MRGn}), in the
scheme $g^2 \to g^2 (1+b_1/b_0 g^2)^{-1}$, and in the original scheme
respectively [in brackets] up to
fourth order. The third and fourth orders include information from the exact
$1/N^2$ RG dependence.} \end{table}
%%%%%%%%%%%%%%%%%%%%%%%%%
For completeness, we also studied in Table \ref{padetabN2} 
similar PA results, but obtained
when truncating to $1/N^2$ the perturbative coefficients beyond second
(perturbative) order, and thus consistently compared to the $1/N^2$ expansion
of the exact mass gap.  (This is motivated from the fact that such coefficients
were generated from the $1/N^2$ information\cite{GraceyN2} exact for the RG
functions, but the higher order dependence $O(1/N^3)$, artificially
generated by our algebraic procedure (detailed in Appendix \ref{ApN2}), is
clearly not the exact one). The numerical results are very good, especially
in the $\MS$ scheme, where it almost indicates a (numerical) 
convergence as the perturbative order increases. \\
%%%%%%%%%%%%%%%%
\begin{table}[htb] \centering
\begin{tabular}{||l||l|l|l|l||}  \hline
$N$ ($O(2N)$ model) & $1/N^2$ ``exact"  & order 2  &  order 3  
   &  order 4       \\ \hline 
2   & 1.7293   &     &  1.781 [1.761] &  1.778 [1.757]
\\ \hline 
3  & 1.4246  &   & 1.438  [1.4262] & 1.434 [1.4236]\\ \hline 
5 & 1.2252  &  & 1.254 [1.2264] & 1.251
[1.2244]\\ \hline
8 & 1.1304  &  & 1.160  [1.1322] &  1.1575  [1.1307]\\
\hline
$\infty$  & 1 &  & $\simeq$ 1 & $\simeq$ 1\\
\hline
\end{tabular}
\caption{\label{padetabN2}$M^P/\Lam$ at $1/N^2$ from a Pad\'e approximant
$P[2,3]$, in the scheme choice (\ref{npscheme}), optimized with
respect to the scale parameter $a$. Only the $1/N^2$ expansion
of the $n \ge 3$ perturbative coefficients is taken into account. The
(unoptimized) $\overline{MS}$ values (i.e. $a=1$) are indicated in brackets.}
\label{tabpadN2}
\end{table}
%%%%%%%%%%%%%%%%%%%%%%%%%
Though the previous numerical behaviour may be 
considered quite satisfactory, the choice of PA is not
unique, and different PA may indeed give quite different values
of the mass gap\footnote{A detailed systematic
analysis of various orders of PA is performed in \cite{damthese}.}. 
For instance, when the
order increases, one can obtain in some cases more than one extrema
with respect to the scale dependence (\ref{padscale}). Moreover, 
as already mentioned the scheme is not completely fixed 
by our procedure, 
and when considering the full scale and scheme
resulting arbitrariness, as well as all possible PA at a given
perturbative order, one eventually finds numerous optima, so that it is
difficult to decide which one is closest to the exact result 
without prior knowledge
of the latter. This reflects the fact that our PA, constructed from the
standard perturbative expansion, are still not able to get rid 
completely of the usual
large freedom due to the scale and scheme arbitrariness in renormalizable
theories. On the other hand, calculating directly (i.e. without optimization)
in the $\MS$ scheme (in which incidentally the exact mass gap
results\cite{FNW} were obtained), we observe from Tables \ref{padtab},
\ref{padtab2} and \ref{padetabN2} that the PA $P_{2,3}$ results, thus depending
on five parameters, appear optimal. Clearly, PA of lowest orders are
inappropriate to correctly match the complete information from perturbative
expansion beyond second order. Second, and quite interestingly, it happens
that the higher order PA almost systematically have poles at $u >0$, rendering
the integration in Eq.~(\ref{padenum}) ill-defined  (and therefore numerically
unstable). In fact,  as explained in Appendix B, 
the expected factorial behaviour at large perturbative orders
is rooted in our PA construction,
which
involves
at perturbative order $n$ a contour integral of the form:  
\be
\ds
I^p_c \equiv \frac{1}{2\pi\,i} \;\oint dv \;e^v \ln^p v \;
\raisebox{-0.4cm}{~\shortstack{ $\sim$ \\ $p\to\infty$}} p!
\ee
which is, not surprisingly, very similar to the standard
renormalon behaviour Eq.~(\ref{irren}) discussed in section 3. 
In other words, though the PA gives a resummation method clearly different from
the Borel method, the resulting integral in Eq.~(\ref{padenum}) 
remarkably shares with the
latter some similar properties, also exhibiting the singularities
of large perturbation orders. 
But, as already mentioned, the choice of PA is not uniquely fixed
e.g. by purely physical considerations. 
All this gives intrinsic limitations to such PA approach,
which make it inadequate to check rigorously numerical convergence at very
high orders, as it does not take advantage of the obtained Borel convergence
properties, discussed in previous sections. 
It should likely be possible to define different PA,
which may better  
exploit the Borel convergence properties,  
avoiding in this way the usual factorial divergences, though 
we refrained to try such analysis here. 
It is in fact simpler to
consider other kinds of 
numerical approximations, more directly related to
the good convergence properties of Eq.~(\ref{contour1})--(\ref{BInl}), as we
examine next.
%%%%%%%%%%%%%%%%%%%%%%%%%%%%%%%%%%%%%
%%%%%%%%%%%%%%%%%%%%%%%%%
%  
\subsection{Borel Resummation results}
Finally in this sub-section we investigate to some extent 
the Borel convergent resummation constructed in sections 5 and 6.
We first show in Table \ref{tabBsum} the  
$M^P/\Lam$ values obtained in the $\hat m \to 0$ limit of the (second RG order)
direct Borel sum expression Eqs.~(\ref{directBS2}), (\ref{RG2Bsum}). We give
here the absolute value $|M^P|/\Lam$ due to the fact that,
at second RG order, Eq.~(\ref{RG2Bsum}) picks up an imaginary part, $\sim
(-1)^{-C}$. (This does not indicate an ambiguity of the
Borel resummation: it is simply due to the branch cut $Re[F] \le -C$
already present within the pure RG dependence, see e.g. Eq.~(\ref{MRGF})).  
As one can see, the results are not very good, except for the
general trend that $M^P/\Lam \to 1$ as $N \to \infty$. 
This is not much
surprising since, as already mentioned, 
Eq.~(\ref{RG2Bsum}) is based only on the large order behaviour of the
perturbative coefficients. We may eventually expect better results
if we could include in a consistent manner the exact $N$ dependence of
the first few perturbative order coefficients rather than their asymptotic
approximations.  This appears qualitatively similar to the simplest oscillator
case, where  the direct Borel sum is known to be a poor approximation to the
exact energy levels.\\  Second, we consider
the non-linear DE-VIP resummation method, explained
in detail in section 6.2, with the resulting 
Eq.~(\ref{nlsum}) for the large order behaviour, but now truncating
this asymptotic behaviour and taking the first two exactly known perturbative
coefficients in Eq.~(\ref{d1d2GN}).  Then, optimizing with respect to the new 
mass parameter $m^\prime$, we obtain the results shown in table \ref{tabnl}.
As one can see, those results, which now largely exploit the good
convergence properties of Eqs.~(\ref{nlresum}),(\ref{BInl}) (and the
true $N$ dependence of the first perturbative coefficients)
are in more reasonable agreement with the exact mass gap, than the
naive optimization results of section 7.1, or the Borel sum results
in Table \ref{tabBsum}. But they
are not as good as the PA results in section 7.2. As already mentioned, this is
likely due to the fact that, even if our alternative series are formally Borel
convergent, the use of such Borel summations is numerically limited. 
Still it is quite satisfactory that within this non-linear and
(formally) convergent alternative expansion, there always exist optima, 
reasonably close to the exact results, which
are moreover relatively flat, and stable against inclusion of higher 
perturbative orders. \\ 
% 
%%%%%%%%%%%%%%%%
\begin{table}[htb] \centering
\begin{tabular}{|c||c|c|}
\hline
N ($O(2N)$)& $M^{exact}_P/\Lam$ & $|M_P|(\hat m=0)/\Lam$ \\
\hline
\hline
2 & 1.8604 & 3.66 \\ 
\hline
3 & 1.48185 & 1.69 \\
\hline
4 &  1.3186 & 1.40 \\
\hline
5 & 1.23668 & 1.27 \\
\hline
8 & 1.1330 & 1.138 \\
\hline
$\infty$ & 1 & $\sim $ 1. \\
\hline
\end{tabular}
\caption{\label{tabBsum} 
$M^P/\Lam$ values as obtained in the massless limit of the second RG order
direct Borel sum expression Eq.~(\ref{RG2Bsum}).} \end{table}
%%%%%%%%%%%%%%%%%%%%%%%%%
% 
%%%%%%%%%%%%%%%%
\begin{table}[htb] \centering
\begin{tabular}{|c||c|c|}
\hline
N ($O(2N)$)& $M^{exact}_P/\Lam$ & $M^{opt}_P/\Lam$ \\
\hline
\hline
2 & 1.8604 & 1.820 \\ 
\hline
3 & 1.48185 & 1.375 \\
\hline
4 &  1.3186 & 1.249 \\
\hline
5 & 1.23668 & 1.187 \\
\hline
8 & 1.1330 & 1.108 \\
\hline
$\infty$ & 1 & $\sim $ 1. \\
\hline
\end{tabular}
\caption{\label{tabnl} Optimized values 
of $M^P/\Lam$ with respect to $m^\prime$ within the non-linear DE-VIP method
Eq.~(\ref{nlsum}), replacing the first two perturbative coefficients with
their exact expressions.}
\end{table}
%%%%%%%%%%%%%%%%%%%%%%%%%
%
\newpage
\section{Conclusion and prospects}
In this paper, we have re-analysed in some detail 
a previous construction
where an alternative perturbation expansion, based on  
a physically motivated self-consistent RG mass solution,
can be Borel convergent in a range of the expansion parameter relevant 
for the massless limit of physical quantities
in asymptotically free theories. 
The perturbative infrared renormalon ambiguities of an AFT, usually 
preventing unambiguous
resummation of the standard perturbative expansion,  
are expected to disappear 
(or more precisely to cancel out with non-perturbative contributions) 
in truly non-perturbative
calculations.
However, such explicit cancellations 
are generally not possible to work out explicitly, except in some particular
2-D models and/or approximations, where exact non-perturbative results are
known, e.g., at the next-to-leading $1/N$
order\cite{David,renormalons,Benbraki,KRcancel}. In contrast, 
one of our main result is that the alternative expansion 
in $1/F$ near the relevant massless limit of the AFT can exist
in two modes, thanks to the infrared properties of 
$F$: \\
i) in the standard mode, corresponding to real $F>0$, and
matching the usual perturbative expansion for $\hat m \gg \Lam$, the
perturbative expansion alone has to be  necessarily completed
as usual with ``non-perturbative" power corrections. This is illustrated 
explicitly with the exact $1/N$ calculation of the GN mass gap\cite{KRcancel},
as discussed in section 3. The net result after cancellations,
Eq.~(\ref{genuine}), is an expression of the mass gap which depends neither on
``perturbative" nor ``non-perturbative" contributions. \\
ii) In the ``alternative" expansion mode, corresponding to 
 $F<0$ in the simplest case (first RG order) where $A_0=1$, 
it gives the sign-alternation in perturbative
coefficients required for Borel convergence. More generally, in an arbitrary
RS, $F$ has complex branches near the massless limit $\hat m =0$, so that
the usual infrared renormalon singularities in the Borel plane are moved away
from the real positive axis range of Borel integration. In this alternative
mode the perturbative series is {\em directly} Borel summable and
non-ambiguous. There is, therefore, no explicit non-perturbative (power
correction) contributions needed in principle in this mode. 
We emphasize that these are generic results for a renormalizable AFT, 
encoded in the basic RG properties
of the implicit function $F$. It should be stressed, however, that our
construction does not solve in general the non Borel summability of
AFT for {\em arbitrary} perturbative expansions in $g(p^2)$. Rather,
it is simply that the alternative expansion relevant for the mass gap,
that is for fixed $p^2$
(or similar on-shell Green functions relevant to the massless limit),
is performed in a neighborhood of $\hat m =0$, i.e. $F=0$ or $F=-C$,
which is away from the usual singularities in the (complex)
coupling $g(p^2)$\cite{thooft,renormalons}. \\

In a second stage, we performed a $\delta$-expansion
(variationally improved perturbation), combined with the previous alternative
expansion. The latter reorganization of
perturbative expansions in AFT makes those particularly convenient for
a $\delta$-expansion approach, since they are much
more similar to the oscillator energy levels expansion, exhibiting a
dependence on $\hat m_v/\Lam$ (Eq.~(\ref{Fexp})) which is power-like
(rather than log-like) for a sufficiently small mass $\hat m_v$.  
This also explains intuitively the peculiar damping mechanism of
factorial divergences, obtained when the
trial parameter $\hat m_v/\Lam$ is order-dependently rescaled, 
in analogy with
similar results\cite{deltaconv,deltac} for quantum mechanics. Yet, unlike the
oscillator the sole rescaling of the mass is insufficient (in the linear
$\delta$ expansion) to obtain a readily convergent series, due to the
reminiscence of the RG dependence in $\hat m_v/\Lam$, making the resulting
power series in $\hat m_v/\Lam$ much more involved than the corresponding
oscillator ones. (Intuitively also from pure dimensional analysis, a
difference is that the oscillator energy level expansions in $g/m^3$
makes it possible to rescale the mass, cf. Eq.~(\ref{rescale}), such that it
can  overcompensate the perturbative factorial behaviour
for\cite{deltaconv,deltac} $1/3 <\gamma <1/2$, thus giving a
reorganized series with an infinite convergence radius. While for a
renormalizable $D \ge 2$ field theory, 
it appears only possible, at least within our approach,
to just compensate exactly the perturbative factorial
behaviour, while still being compatible with the massless limit.)   
But, when combined with Borel resummation, the usual
infrared renormalon singularities at finite $t_0 >0$ in the Borel plane are
rejected towards $t_0 \to +\infty$, as a consequence of the damping of
renormalon factorials from the $\delta$-expansion. Still, for real $F>0$ the Borel integral
does not converge at $t \to \infty$, while convergence can be obtained again
in the range $F <0$ or complex. In this case, the $\delta$-expansion combined
with the infrared properties of the $1/F$ expansion can lead to an improved
(fastly Borel convergent) expansion. We stress also that the linear DE-VIP
taking the form (\ref{contour1}), and  (\ref{BIvar})  when combined with the
Borel method, is only one among various similar resummation means. In fact the
improved convergence properties do not depend on the detailed properties of
the contour integrals here considered, e.g. Eq.~(\ref{contour1}) [though those 
have the advantage of giving rather simple and tractable
expressions in the massless limit and for Borel transforms
Eqs.~(\ref{BIvar})]: a non-linear version of the DE-VIP expansion
is also possible, as we investigated here, Eq.~(\ref{BInl}),
which can even lead to a {\em directly} convergent series. 
More generally,  performing a ``brute force"
$\delta$-expansion on e.g. the mass gap Eq.~(\ref{MRGn}), and 
rescaling the trial mass $m_v$ according to (\ref{rescale}),
replaces Eq.~(\ref{Mpolesum}) and subsequent results 
with more complicated series, but having
similar asymptotic and (Borel) convergence properties (see Appendix E). 

We have then performed a rather detailed numerical application of this
construction for the $O(N)$ GN model, taking the mass gap (exactly know
for arbitrary $N$ from other non-perturbative approaches specific of integrable
models) as a test
of our method. Though the construction is quite general, it is restricted in
practice by the lack of knowledge beyond the few first orders of the genuine
perturbative coefficients in most $D \ge 2$ field theories. In order to take
into account as much as possible these uncertainties,   
we considered various different approximations: \\
--direct numerical ``PMS"
optimization with respect to the trial mass 
of the alternative $1/F$ expansion form, or the $\delta$-expansion form of the
mass gap, with the exact first two perturbative coefficients;\\
--Pad\'e
approximant resummations, including higher order RG information at the $1/N^2$
level; \\
--``Borel-inspired" resummation, assuming ony the leading asymptotic
behaviour of the perturbative coefficients.\\ 
We obtained numerical results which are more or less consistent
with each others, and showing in some cases a very good agreement with the
exact results, at the few percent level or less, 
even for relatively small $N$ values\footnote{Recently
another variant of the PMS optimization approach, 
so-called ``source inversion" method\cite{AcoVer}, 
also gave reasonably good numerical results for the GN
model. This latter approach, which shares some similarities
with our method, does not address, however, the important problem 
of the large perturbative order behaviour.}. 
The best results are obtained from a certain class of Pad\'e approximants. 
These PA exhibit by construction some of the analytical properties of the 
Borel sums, at least qualitatively,  
but can more conveniently incorporate the exact (non-asymptotic) lowest
orders perturbative dependence as well as systematic higher order corrections.
The optimization results are also in general better than those obtained from
direct Borel sum expressions. 
(This is to some extent analogous with the Borel summable
oscillator case, where the direct Borel sum 
is of not much practical use for a numerical determination
of the energy levels, more accurately obtained either by Pad\'e-Borel
resummations\cite{aoBsum}, or the ODM \cite{odm} and related
optimized delta-expansion \cite{deltaconv,deltac} methods). We
cannot conclude from our numerical analysis, however, that a definite and
rapid numerical convergence is obtained in field theories, in contrast with
similar analysis performed for the oscillator energy levels. We think
this may be mostly due to the large differences in renormalizable field
theories between the true perturbative low order coefficients and their
asymptotic behaviour,  together with the fact that there are some practical
limitations preventing to perform our analysis to arbitrary high orders,
unlike the oscillator case. Nevertheless, we are confident that it should
exist a better numerical exploitation of our construction, able to
approach arbitrarily close to the exact mass gap from purely perturbative
information, by  taking advantage in a
more efficient way of the formal Borel convergence 
properties here discussed.\\   
Finally, since the construction is valid a priori for any AFT,
and can rely only on the perturbative information which is available
in many models, we argue  that such a summation recipe can
provide a well-defined basis 
to estimate more precisely some of the
\CSB order parameters in more complicated, 4-D theories like
QCD or other 4-D
models with a nontrivial \CSB structure.  
In QCD, 
rather than the quark pole mass which is ill-defined due to the confinement,
the quantities of much
interest are certain (gauge-invariant) condensate operators
which are the order parameters
of the (dynamically broken) chiral symmetry. These
can be inferred in the chiral limit $\hat m\to 0$ similarly
from this alternative expansion\cite{qcd2}. 
A detailed
investigation in QCD is however beyond the scope of the present paper.\\

\begin{acknowledgments}
We thank Andr\'e Neveu for valuable discussions.
We are also grateful to G\'erard Mennessier for useful comments on 
Borel summation methods. 
\end{acknowledgments}
%%%%%%%%%%%%%%%%%%%%%%%%%% 
\appendix
\section{Formal RG resummation at arbitrary orders}\label{RGall}
\setcounter{equation}{0}
In a scheme where $b_i, \gamma_i \neq 0$ for $i \ge 2$, one can generalize
formally expression (\ref{MRGn}). We obtain
\bea 
 M^P (\hat m) 
 = 2^{-C} \frac{\hat m}  
{F^A_2 [C + F_2]^B}   
 {\sum^\infty_{n=0} \frac{d_n}{(2b_0 F_2)^n}}, 
\label{MRGN} 
\eea 
where
by construction we kept for convenience  a form similar to the second order,
by integrating the basic RG equation (\ref{runmass}) 
separating explicitly  the $b_0$, $b_1$, $\gamma_0$, $\gamma_1$ dependence
from the higher order coefficients, that are integrated perturbatively, 
as they give a simple polynomial dependence. 
Thus in (\ref{MRGN}) we explicitly separate a {\em resummed}
RG dependence, the term $F^{-A}_2 (C+F_2)^{-B}$, from the remnant which
is systematically expanded in powers of $F_2^{-n}$, where $F_2$ is given
by the exact second order relation in (\ref{F2def}): the form
of (\ref{MRGn}) as a function of $F$ is not unique at arbitrary orders:
the important point with (\ref{MRGn}) is that all higher
order RG dependence on $b_n$, $\gamma_n$ RG coefficients can be put into
the form of contributions to the coefficients $d_n$ of $F_2^{-n}$,
as explicitly shown below.
Accordingly we can write 
\be
d_n \equiv c_n + r_n
\label{dn}
\ee 
to distinguish contributions from 
the true perturbative part, $c_n$ (i.e. the non-log part of the n-loop
perturbative graph) from $r_n$, the contribution for $n \ge 2$ originating from
higher order RG-dependence (in a scheme different from 
Eq.~(\ref{betuni}),(\ref{gamuni})) due to the
re-expansion in powers of $1/F$ implicit in (\ref{MRGn}) or (\ref{MRGN}).\\
At arbitrary order $n$, $F(\hat m)$, $\Lam$ and the invariant mass $\hat m$
are different
from the corresponding second RG order expressions.
Their expressions can be again formally derived order by order in
perturbation. For instance
\beq 
F(\hat m/\Lam) \equiv F_2
+\sum_{n \ge 1} \alpha_n/F_2^n 
+\sum_{n \ge 2} \delta_n/F_2^n 
\label{Fndef} 
\eeq
\be
\hat m = \hat m_2 \exp[-\frac{\gamma_0}{2b_0} \sum_{n \ge 2} \delta_n
g^{2n}]
\label{mndef}
\ee
where 
\bea
&\alpha_1 &= -\frac{b_2}{2b^2_0}\;,\;\;
\alpha_2 =  \frac{b_1 b_2}{2b^3_0} - \frac{b_3}{4b^2_0}\;,\cdots \nn \\
&\delta_2 &= \frac{b_0\gamma_2 -b_2\gamma_0 }{8b^3_0\gamma_0}\;,\cdots
\eea
and $\hat m_2$ in 
Eq.~(\ref{mndef}) designates the second RG order
scale invariant mass, defined in Eq. (\ref{m2def}).
Similarly,  $\Lam$ at arbitrary RG orders $n$ involves  perturbative
corrections with   respect to the (universal scheme) expression
in (\ref{Lamdef}): 
\be
\ds \Lam = \mu\;\exp({-\frac{1}{\ds 2\,b_0\,g^2(\mu)}})\; 
[b_0\,g^2(\mu)]^{-C}\;[1+\frac{b_1}{b_0}g^2]^C
\;[1+\sum_{i \ge1} \lambda_i g^{2\,i}(\mu)\:]
\ee
with
\be
\lambda_1 = -\frac{b_2}{2b^2_0}\;,\;\;
\lambda_2 = 
 -\frac{2b^2_0 b_3 -4b_0 b_1 b_2 - b^2_2}{8 b^4_0}\;,\cdots
\ee
All the above relations
expresses consistently the fact that $\Lam$ is scale invariant 
and $\hat m$ is scale and scheme invariant at a given $n$ order. 
Moreover, the connection between the exact second order part $F_2$ 
of $F$, which is not
perturbatively expanded, and the exact second order part $\hat m_2$ of
$\hat m$, 
is such that one still has the property
\be
\hat m \;F^{-A} (C+F)^{-B} \to const.\; 2^{-C}\; \Lam \;[1+ {\cal O}(\frac{\hat
m}{\Lam})\:] \label{M2Lam}
\ee 
when $\hat m \to 0$, where $A= \gamma_1/(2b_1)$, thus maintaining the
infrared properties of the massless limit discussed at first and second RG
orders in section 2. 
%%%%%%%%%%%%%%%%%%%%%%%%%%%%%%%%%%%%%%%%%
\section{Oscillator energy levels and Borel summability}\label{ao}
\setcounter{equation}{0}
In this appendix we briefly elaborate on a perhaps unusual picture of the
anharmonic oscillator energy levels, in order to illustrate in a simpler case
the existence of two different perturbative expansion modes, with one mode
needing additional non-perturbative corrections while the other mode is
directly Borel summable.\\
We recall that the oscillator is described by a $D=1$, $g\phi^4$ 
massive scalar field theory\cite{ao}:
\be
{\cal L}_{ao} = \frac{1}{2}(\partial_t \phi)^2 -\frac{m^2}{2} \phi^2
-\frac{g}{4!}\phi^4 
\label{Lao}
\ee   
The energy levels have a perturbative
expansion\footnote{We shall
only consider the ground state energy level to simplify,  which is sufficient
for our purpose.}
\be
E_0 \sim |m|\;\sum_n a_n (\frac{g}{|m|^3})^n
\label{E0exp}
\ee
where the coefficients can be calculated to arbitrary orders, and have
the well-known asymptotic behaviour\cite{ao,zinn}:
\be
a_n \sim -(-1)^n (\frac{6}{\pi^3})^{1/2} (\frac{3}{4})^n \Gamma[n+1/2]\;
(1+{\cal O}(\frac{1}{n}))\;.
\label{aoco}
\ee
Because of the sign-alternation of the coefficients in Eq.~(\ref{aoco}), the
series
is Borel-summable\cite{aoBsum}. Consequently,
there are no additional ``non-perturbative" contributions, 
and this series can be uniquely Borel-resummed to represent
the (real part of) the oscillator ground-state energy.
Explicitly, the Borel sum reads:
\be
\tilde E_0 (g) \sim - \frac{(6)^{1/2}}{\pi}\; \frac{|m|^3}{g}\;
\int^\infty_0 dt e^{-t\,\frac{|m|^3}{g}}\;(1+\frac{3}{4}t)^{-1/2}
\label{aoBsum}
\ee 
Notice that the result Eq.~(\ref{aoco}), originally
obtained from a WKB calculation\cite{ao}, was in fact later shown to be
derivable from an instanton-based calculation, the
vacuum being unstable for $g <0$ with a corresponding tunneling process. More
precisely, a classical calculation gives\cite{zinn} for the instanton
contribution to the imaginary part of the ground-state energy:
\be
Im\, E_0 \sim \frac{4}{\sqrt{2\pi}}\;
e^{\frac{4|m|^3}{3g}}\;(\frac{|m|^3}{-g})^{1/2} 
\label{aoinst}
\ee
where $g <0$. \\
Now let us assume that our only knowledge would consist of the 
purely perturbative information, namely the expansion in Eq.~(\ref{E0exp}),
and consider formally changing the sign of the
coupling there: $g \to -g$. Obviously this cancels
the sign-alternation of the coefficients,
so that the corresponding new series is no longer Borel summable:
more precisely instead of Eq.~(\ref{aoBsum}) one obtains an integral
\be
\tilde E_0 (g) \sim \frac{(6)^{1/2}}{\pi}\; \frac{|m|^3}{|g|}\;
\int^\infty_0 dt e^{-t\,\frac{|m|^3}{|g|}}\;(1-\frac{3}{4}t)^{-1/2}
\label{aoBsum2}
\ee 
which is ill-defined due to the cut at $t=+4/3$ on the integration range.
Nevertheless, we can evaluate the ambiguity that this implies: by defining
the ambiguity by the (half) difference of the two possible contours avoiding
the cut by above (resp. by below), a straightforward contour calculation gives
an ambiguity: 
\be
\delta E_0 \sim  \frac{4\,i}{\sqrt{2\pi}}\;
e^{-\frac{4|m|^3}{3|g|}}\;(\frac{|m|^3}{|g|})^{1/2} 
\label{aoambig}
\ee
which accordingly is to be interpreted as an additional
non-perturbative contribution needed in this unconventional picture. 
Not suprisingly, one
recovers in fact consistently the 
non-perturbative instanton contribution
Eq.~(\ref{aoinst}), with $g<0$. (NB we recall
that actually, it was the instanton solution Eq.~(\ref{aoinst}) that lead to
{\em derive}\cite{zinn} the asymptotic behaviour in Eq.~(\ref{aoco}), so that
Eq.~(\ref{aoambig}) is nothing but a consistency check,
though relevant to our argument.) 
 Thus, the two perturbative expansion modes,
the directly Borel summable one with $g>0$ or $g<0$ respectively, can be
both consistent, provided one correctly identifies the necessary additional
non-perturbative contributions in the $g<0$ case.
%%%%%%%%%%%%%%%%%%%%%%%%%%%%%%%%%%%%%%%%%
\section{Pad\'e approximants and chiral limit}\label{Appade}
\setcounter{equation}{0}
In this appendix we give some technical details on the construction
of the Pad\'e approximants (PA) used in sec.\ref{pade}. 
Similar PA were used in ref.~\cite{gn2}.
As mentioned, the aim
is first to have a well-defined (finite) massless limit,
$F\to 0$, $\hat m/\Lam \to 0$, when including the purely
perturbative expansion series. 
Eq.~(\ref{F2def}) suggests that for the contour integrand
e.g. Eq.~(\ref{contour1}), where $F\equiv F[m^{\prime\prime}\:v]$, one
reintroduces instead of $m^{\prime\prime} \equiv \hat m/\Lam$ an (inverse)
``effective coupling" variable $u$ defined by  \beq
 u^A e^u (C+u)^{B-C} = m^{\prime\prime} .
\label{udef}
\eeq 
In the usual perturbative regime, $m^{\prime\prime}\to \infty$, in terms of $u$, the $\ln \ln
\ldots \ln m^{\prime\prime}$ disentangle, and Eq.~(\ref{F2def}) admits an
asymptotic expansion for $u \rightarrow \infty$:  
\bea
& F[m^{\prime\prime}\:v] & \sim  u \; \left[1 +\frac{\ln v}{u}  -\frac{(A+B-C)}{u}
\ln\frac{F}{u} - \frac{(B-C)}{u} \ln[\frac{(1+C/F)}{(1+c/u)}] \right] \nn \\
&  & \sim  u \; \left[1 +\frac{\ln v}{u}  -\frac{(A+B-C)\;\ln v}{u^2}
+{\cal O}(u^{-3})\right] 
\label{Fpade}
\eea
where the higher order terms are polynomial in $\ln v$, and evidently depends
on the RG coefficients $A, B, C$. Using Eqs.~(\ref{udef}), (\ref{Fpade}) within
the contour integral Eq.~(\ref{contour1}), around the cut at $Re[v] <0$ defines
the mass gap as an asymptotic
expansion: 
\bea
\ds
&\frac{M^P}{\Lam} &= const.\; e^u \;u^{-(C+1)} \times \nn \\
& & \left[A_0 +\frac{1}{u}((A_0 +d_1 +d_1 A_0 -A_0(A_0-C) +\gamma_E
A_0(A_0+1)) +{\cal O}(u^{-2})\;\right]
\label{Mpade}
\eea
where $A_0 \equiv A+B$, and $d_i$ are the original perturbative coefficients
of Eq.~(\ref{MRGn}) in a given scheme, including 
eventually the $1/N^2$ RG-dependence (see Appendix \ref{ApN2}). The
overall constant includes any  constants from e.g. RG dependence (such as
typically the $2^{-C}$ factor in Eq.~(\ref{contour1})),
or depending eventually on the
choice of renormalization scheme. In order to derive perturbative expansion
such as Eq.~(\ref{Mpade}) at 
arbitrary higher orders $u^{-p}$, $M^P/\Lam$  has to be
systematically expanded using
\be
\ds
I^p_c \equiv \frac{1}{2\pi\,i}\; \oint dv\; e^v \ln^p v = \frac{i}{\pi}
\sum^{p-2}_{j=1}\; (^p _j) \;(i \pi)^j \;I_{p-j} 
\label{contlog}
\ee
where $(^p _j)$ are the binomial coefficients and
\be
\ds
I_n \equiv \int^\infty_0 dv \;e^{-v} \ln^n |v|
\label{reallog}
\ee
can be evaluated exactly in terms of Euler's constant $\gamma_E$ and the
Riemann zeta functions $\zeta[n]$. We give in table \ref{Iintres} the first few
$p$ orders for those integrals. Note that the contour integral results $I^p_c$
in (\ref{contlog}) are real.
\begin{table}
\begin{center}
\begin{tabular}{|c|c|c|}
\hline
p  & $I_p$  & $I^p_c$ (contour)\\
 & &  \\
\hline
0  & 1 & 0 \\
 & &  \\
\hline
1  & $-\gamma_E \simeq -0.577216$ & 1 \\
 & &  \\
\hline
2  & $\gamma^2_E +\zeta[2]\simeq   1.97811$ &   1.15443 \\
 & &  \\
\hline
3  &  $ -\gamma^3_E -\frac{\pi}{2}\gamma_E -2\zeta[3] \simeq  -5.44487$
&  3.93527 \\ 
 & &  \\
\hline
4  & $\gamma^4_E +\pi^2 \:\gamma^2_E +\frac{3\pi^4}{20} +8 \gamma_E \zeta[3]
\simeq  23.5615$ &  -1.00806 \\ 
 & &  \\
\hline
5  &   $\cdots 
\simeq -117.839 $ &  -19.9846\\
 & &  \\
\hline
\end{tabular}
\vspace{5mm}
\end{center}
\caption{\label{Iintres} 
Results of integrals Eqs.~(\ref{reallog}), (\ref{contlog})
for the first few $p$ values.} 
\end{table}
%%%%%%%%%%%%%%%
We remark that the basic integral $I_p$ behaves as $\sim (-1)^p \: p!$ as
$p$ increases, and similarly the final contour integral $I^p_c$ behaves
as $\sim (p-1)!$, with no definite sign. \\
Taking now the derivative with respect to $u$ of the logarithm of this
expression for $M^P/\Lam$ yields the final power series in $1/u$
\be
\frac{\partial \ln [ \frac{M^P}{\Lam} ]
}{\partial u }  \simeq 
1 \; - \; \frac{C+1}{u} \, +{\cal O}(u^{-2})
\label{Mderpade}
\ee
suitable for PA analysis.
Similarly to \cite{gn2}, we thus consider
PA $P_{[p,q]}$ as rational fractions of polynomials:
\be
\ds
P_{[p,q]}(u) \equiv \frac{1+\sum^p_{n=1} s_n/u^n}
{1+\sum^q_{n=1} t_n/u^n}
\label{padegen}
\ee 
with coefficients $s_n$, $t_n$, uniquely fixed from requiring
that the perturbative expansion, for $u\to\infty$, matches
the perturbation series obtained from Eq.~(\ref{Mderpade}), {\em and} that
the massless limit $u\to 0$ is finite. Actually since our PA
are constructed from  $\partial\,\ln[M^P/\Lam]/\partial u$, one further
step is needed to recover the mass gap itself. We get more precisely
\be
\ln \frac{M^P}{\Lam}(u\to 0) =
\ln \frac{M^P}{\Lam}(u\to \infty) 
-\int^\infty_0 du \;P_{[p,q]}(u) 
\ee
where the  initial condition 
\be
\ln \frac{M^P}{\Lam}(u\to \infty) = \ln[const. A_0] +u -(C+1)\ln u +
{\cal O}(1/u)
\ee
is uniquely fixed because it is entirely determined by the first order of
perturbation theory. Its (perturbative) divergences for $u\to\infty$ are
exactly compensated by the corresponding ones from the first terms of the
Pad\'e approximants, and we finally obtain
\be
\frac{M^P}{\Lam} = const. A_0 \exp \left[
1- \int^1_0 du \;P_{[p,q]}(u) - \int^\infty_1 du \;[P_{[p,q]}(u)-1
+\frac{C+1}{u} ] \right]
\ee
which is used in the numerical analysis in section \ref{pade}.
%%%%%%%%%%%%%%%%%%%%%%%%%%%%%%%%%%%%%
\section{Finite $1/N^2$ corrections from scheme
change}\label{ApN2} 
\setcounter{equation}{0}
In this appendix we give relevant formulas for the RS change
which defines the (approximated) $1/N^2$ corrections to the 
finite perturbative coefficients of arbitrary orders. These are used for
the numerical analysis of Pad\'e approximants in section \ref{pade}.\\
The first step is to extract from  the results of ref.~\cite{GraceyN2}
the perturbative expansions of the RG functions $\beta(g)$ and $\gamma_m(g)$ 
to arbitrary orders in the $\MS$ scheme,
which serves as the basis of our RS change. In \cite{GraceyN2}, by studying the
theory near the (perturbative) fixed point $g_c$ of the 
$\epsilon$--dependent beta function:
\be
\beta(g,\epsilon) \equiv \epsilon +\sum_{i=0} (-2\,b_i) (g^2)^{2+i}
\label{beteps}
\ee  
were derived the
exact $1/N^2$ expressions, as function of $\epsilon \equiv D-2$: 
\be
\beta^{'}(g_c) \equiv \frac{\partial \beta}{\partial g^2}(g_c) 
= -2 [\mu -1 -(2\mu-1) \frac{\eta_1}{N} +\frac{\lambda_2}{N^2}]
\label{betprime}
\ee
and
\be
\gamma_m(g_c) = D-2 +\frac{\eta_1+\chi_1}{N}+
\frac{\eta_2+\chi_2}{N^2}\;.
\label{gamprime}
\ee
In (\ref{betprime}) and (\ref{gamprime}) $\mu \equiv (D+\epsilon)/2$ and the 
coefficients $\eta_i$, $\lambda_i$ and $\chi_i$ are given analytically in terms
of the Gamma function $\Gamma[z]$ and its first two derivatives, $\Psi[z]$ and
$\Psi^{'}[z]$. We refer to Ref. \cite{GraceyN2} for details on this derivation
and explicit expressions of $\eta_i$, $\lambda_i$, $\chi_i$ 
which we essentially follow except for
slight differences of conventions e.g. in the definition (\ref{beteps}).\\
The first three-loop orders $b_i$, $\gamma_i$ up to $i=2$ are 
known for arbitrary $N$ in the $\MS$ scheme\cite{Gracey3loop}. Next,
by simply comparing the expansion of $\beta^{'}(g_c)$ to arbitrary orders in
$\epsilon$, using Eq.~(\ref{beteps}):
\be
\beta^{'}(g_c)[\epsilon] = -\epsilon -\frac{b_1}{2b^2_0} \epsilon^2
 +\frac{b^2_1-b_0b_2}{2b^4_0}\epsilon^3 +\cdots
\label{betepsexp}
\ee
(where $b_n$ appears first at order $\epsilon^{n+1}$)
with the same $\epsilon$ expansion obtained from the exact $1/N^2$ expressions
(\ref{betprime}), (\ref{gamprime}), which gives e.g. for $\beta'(g_c)$:
\be
\beta^{'}(g_c)[\epsilon] = -\epsilon +\frac{N+2}{N^2}\epsilon^2
+\frac{N+1}{2N^2}\epsilon^3 
-\frac{47 + 3 N - 101 \psi^{\prime\prime}[1] + 2 \psi^{\prime\prime}[2]}{12N^2} \epsilon^4
+\cdots\;, \label{bpgcex}
\ee
it is straightforward to obtain e.g. for the $n \ge 4$ loops unknown RG
coefficients: 
\bea
b^{\MS}_3 &&= \frac{(2 - N)}{384\pi^4\:N^2} [136 - 196N + 126N^2 -
19N^3 - N^4 \nn \\
&&+\psi^{\prime\prime}[1](- 264 + 396N -198N^2 +33N^3)]
\label{betgc3}
\eea
and
\bea
\gamma^{\MS}_3 &&= \frac{1}{384\pi^4\:N^2} [
176 - 424N + 350N^2 - 128N^3 + 18N^4 + 5N^5 \nn \\
&&+ \psi^{\prime\prime}[1](144 + 
 24N - 276N^2 + 198N^3 - 48N^4 + 3N^5)]
\label{gamgc3}
\eea
where we refrain to give higher orders which involve quite lengthy
expressions\footnote{ All algebraic calculations were performed with
Mathematica\cite{matha}. Note in Eqs.~(\ref{betprime})--(\ref{gamgc3}) that $N$
now refers to the $O(N)$ model.}. 
Next step is to incorporate this exact $1/N^2$ information as corrections to
the 't Hooft scheme mass gap expression (\ref{MRGn}). To this aim, we
exploit the fact that the specific change of scheme implied by passing from
the $\MS$  RG functions Eqs.(\ref{beteps})--(\ref{gamgc3}) to the truncated
ones in the two-loop 't Hooft scheme Eqs.~(\ref{betuni}),(\ref{gamuni}) can
be achieved by appropriate redefinitions of the  coupling $g(\mu)$
(equivalently redefining $1/F$) and redefinitions of the mass $m(\mu)$. The
most general perturbative such RS change (at fixed scale $\mu$):
\bea 
\label{A1B1} 
\ds
g^2 \to \tilde g^2 = g^2 (1+ A_1 g^2 + A_2 g^4 + ... ) \nn \\ 
m \to \tilde m \equiv m\;Z_m(g) =
m (1+ B_1 g^2 + B_2 g^4 + ... )
\eea 
implies from\cite{Collins}
\bea
\ds
&\tilde g\:\tilde \beta(\tilde g) \equiv g\:\beta(g) \;\frac{\partial
\,\tilde g^2 }{\partial g^2}\nn \\ &\tilde \gamma_m(\tilde g) = \gamma_m(g)
-2g\;\beta(g)\;\frac{\partial }{\partial g^2}    \;\ln Z_m(g)
\eea
relations between the RG coefficients in the two schemes\cite{qcd2}
\bea
\tilde \gamma_1 &&=\gamma_1 +2b_0 B_1 -\gamma_0 A_1 \nn \\
\tilde b_2 &&= b_2 -A_1 b_1 +b_0(A_2-A^2_1) \nn  \\
\tilde \gamma_2 &&=\gamma_2 +2b_1 B_1 +2b_0 (2B_2-B^2_1) -2A_1 \tilde \gamma_1 
-\gamma_0 A_2
\label{RSCRG}
\eea
and corresponding changes in the purely perturbative coefficients
of (\ref{MRGn}):
\be
\tilde d_1 = d_1 -B_1\;\;\;, \tilde d_2 = d_2 -\tilde d_1 (A_1+B_1) -(B_2
-\gamma_0 B_1)\;\;\cdots
\label{RSCpert}
\ee
The RS transformation is uniquely fixed, apart from $\gamma_1$, by requiring  
the new scheme to be the above $1/N^2$ $\MS$ Eqs~(\ref{betgc3}),
(\ref{gamgc3}) and the old scheme to be the two-loop truncated RG
coefficients. There is however one missing piece in this construction:
the exact $1/N^2$ dependence above is the one of the RG coefficients, while
the genuine (non-RG) finite parts, are actually unknown for the GN model beyond
two-loops. Thus, in addition to fix uniquely the perturbative contributions
 $\tilde d_n$ in Eq.(\ref{MRGn}) in the new scheme, we assume that these are
negligible beyond two loops, in the original ('t Hooft) scheme,  $d_n \simeq 0$
for $n >2$. Whether this assumption is  a good approximation or
not can only be decided by the numerical analysis, see section 7B.\\

Actually, the brute force RS change as implied by Eqs.(\ref{RSCRG}),
(\ref{RSCpert}) becomes
rapidly algebraically involved as the expansion order increases. A more
efficient method can be devised\cite{damthese}, more convenient to study Pad\'e
approximants of relatively high orders. Noting that the RS transformation is
uniquely fixed, up to $\gamma_1$, by $A_n$ and $B_n$ in Eqs.(\ref{A1B1}),  one
can  interpret the RS
transformation as a one-parameter $\tau$ variation in the RS parameter space,
with ``boundary" conditions $A_n(\tau)$ and $B_n(\tau)$ for the initial and
final scheme, $\tau=0$ and $\tau=1$, respectively.
Using that the bare coupling and mass are RS invariants, one defines 
\bea 
\label{A2} 
\alpha = \frac{dg}{d\tau}\nn \\ 
\delta_m = \frac{dm}{d\tau} 
\label{rscfunc}
\eea 
with perturbative expansion 
\bea
\label{A3} \alpha (g,\tau) &=& -a_0 g^3 - a_1 g^5 - ... \nn \\ 
\delta_m(g,\tau) &=&  D_0 g^2 +D_1 g^4 + ... 
\label{rscfunc2}
\eea 
similarly to the above ordinary RG functions $\beta(g)$ and $\gamma_m(g)$.
While the latter parameterize the variation with the scale $\mu$,
the functions in Eqs.~(\ref{rscfunc}) parameterize the variation in the 
most general RS parameter space (which is $2n+1$
dimensional at perturbative order $n$), 
with $\tau$ playing a role similar to $\ln\mu$, and with $2n$
boundary conditions fixed by the initial and final specified schemes. This
is completely equivalent to the brute force above RS change, but more
convenient, because Eqs~(\ref{rscfunc}), (\ref{rscfunc2})
 can be resummed to some extent, for
the leading and subleading dependence, 
similarly to the well-known RG resummations properties of $\beta(g)$ and
$\gamma_m(g)$. More precisely, imposing for simplicity that the coefficients 
$a_i$ and $D_i$ do not depend on $\tau$, $A_n(\tau)$ and $B_n(\tau)$ are
series expansion in $\tau$ and we also have the following useful relations
between the RS change functions and the ordinary RG functions:
\be
\frac{d\beta}{d\tau} = \frac{d\alpha}{d\mu}\;;\;\;\;\;\; 
 \frac{d\gamma_m}{d\tau} = \frac{d\delta_m}{d\mu}\; 
\ee
which gives by perturbative expansion
\bea
\frac{d}{d\tau}(b_0) &&= 0 \ ;  \nonumber\\
\frac{d}{d\tau}(b_{r-1}) &&= 2\sum_{n=1}^{r-1} (2n-r)a_{r-n-1} b_{n-1}(\tau) \;
\eea
and
\bea
\frac{d}{d\tau}(\gamma_0) &&= 0 \ ;\nonumber\\
\frac{d}{d\tau}(\gamma_{r-1})&&= 
\sum_{n=1}^{r-1} 2n \Big[  a_{r-n-1} \gamma_{n-1}(\tau) -
D_{n-1}b_{r-n-1}(\tau) \Big] \ . 
\eea
Integration of these 
equations leads to
appropriate recurrence relations\cite{damthese}: 
\bea 
\label{recurence3a}
b_{r-1}(\tau) &&= 2\sum_{n=1}^{r-1} (2n-r)\ a_{r-n-1} \int_0^{\tau}ds
b_{n-1}(s) \, \\ 
\label{recurence3b} 
\gamma_{r-1}(\tau) &&=
\sum_{n=1}^{r-1} 2n \Big[  a_{r-n-1} \int_0^{\tau}ds \gamma_{n-1}(s) -
D_{n-1}\int_0^{\tau}ds \gamma_{n-1} b_{r-n-1}(s) \Big] \ . 
\eea 
defining
order by order all the needed RS information.
%
%%%%%%%%%%%%%%%%%%%%%%%%%%%%%%%%%%%%%%%%%%
\section{Standard delta-expansion}\label{standel}
\setcounter{equation}{0}
For completeness we briefly examine here the DE-VIP convergence properties in
our framework by applying to the mass gap expression Eq.~(\ref{MRGn}) the more
standard\cite{delta}--\cite{deltac} order by order $\delta$-expansion 
without our specific contour integral resummation prescription
illustrated in sections 5 and 6. To this aim we
first define the substitution \bea 
& m &\to m_v\;(1+\delta\:\frac{m-m_v}{m_v}) 
\equiv m X (1-\delta \;\beta(X))
\;\;\nn \\
& g^2(\mu)& \to 
\delta\:g^2(\mu)  \label{substit2}
\eea  
where $X\equiv m_v/m$ and $\beta(X) \equiv 1-1/X$,
to apply to
\be
M^P \sim \hat m F^{-A} [1+\frac{1}{2\,b_0}\sum^N_{n=0}
\:\frac{d_n}{F^{n+1}}\;]
\label{mp1}
\ee 
expanded to order $\delta^N$. (Without much loose of generality we again
took for simplicity in Eq.~(\ref{mp1}) the first RG order expression of the
resummed RG dependence). After
straightforward algebra it gives 
\bea
&M^P/\Lambda &\sim
1 +\frac{1}{4\pi\,b_0}\;
\sum^N_{q=1} \sum^{N-q}_{p=0} \sum^{N-q-p}_{r=0} \nn \\
&&\times \;\frac{\Gamma[p+q]
(p+q+A)(q+A)^{p-1}}{A^p\:\Gamma[1+p]\:
}\;(\frac{\Gamma[r+q/A]}{\Gamma[r+1]\Gamma[q/A]})\;
(m^{\prime\prime} X)^{-q/A}\; (\beta(X))^r\;
\label{Mpsum}
\eea
The basic calculation  is similar to the one that lead to Eq.~(\ref{Mpolesum}),
except for the factor $1/\Gamma[1+q/A]$ less (which in (\ref{Mpolesum})
originates from the contour integral), and the third summation on
$r$ in addition, which corresponds to the expansion of $[X (1-\delta
\;\beta(X))]^{-q/A}$. Note also that the delta-expansion is truncated 
at order $N$, and $\delta \to 1$: since  the perturbative  term 
$g^{2n} =(g^2)^{p+q}  \to \delta^{p+q} (g^2)^{p+q}$, it explains
the third summation upper bound $N-q-p$.\\
Note furthermore that here 
the expansion of  $F$ around $0$ is rigorously valid,
as long as  
\be
m X (1- \delta\;\beta(X)) < e^{-A}A^A
\ee
(where the right-hand side corresponds to the convergence radius
of the power expansion form of $F$).\\

It is clear that the terms of the third series,
though considerably complicating the algebra, do not play any role
in the asymptotic/convergence properties.
Looking at  the asymptotic behaviour 
of  (\ref{Mpsum}) it is easily seen, by rescaling
$X \equiv m_v/m$ as
\be
X \to  N^\gamma \;\tilde X\;.
\label{rescale2}
\ee
that the series behaves like the one in (\ref{Mpolesum}),  
for\footnote{After rescaling, even if $X\to\infty$ as $N\to\infty$,
we do not leave the convergence disc of the power expansion form of $F$,
Eq.~(\ref{Fexp}): in fact, $X\to\infty$ since $m \to 0$ with fixed $m_v$, so
one can always choose the arbitrary $m_v$ such that  $m_v (1-
\delta\;\beta(X)) < R_C =e^{-A_0} (A_0)^{A_0}$.}  $A \le \gamma $.\\ 
The series can next be "Borelized" in a way similar to what is described in
section 5 above. It gives more complicated Borel integrals, 
but having similar asymptotic and  Borel convergence
properties, in particular for $Re[m_v]<0$. 
We refrained however
to attempt any numerical analysis based on the complicated series in
Eq.~(\ref{Mpsum}), which appears less convenient for this purpose than
the construction mainly discussed in the rest of the paper.
\end{document}